%% file: main.tex
\documentclass{article}
\usepackage{myarticle}
\newcommand{\rootdir}[1]{./#1} % relative path to root directory 

\newcommand{\citesupp}{}

\title{Bayesian estimation of causal effects from observational categorical data}

\author{Vera Kvisgaard$^1$\thanks{Corresponding Author: verahk@uio.no} \and Johan Pensar$^1$}
\date{%
    $^1$Department of Mathematics, University in Oslo \\[1em]
    \today
}

\begin{document}

\maketitle 

\begin{abstract}
We present a Bayesian procedure for estimation of pairwise intervention effects in a high-dimensional system of categorical variables. We assume that we have observational data generated from an unknown causal Bayesian network for which there are no latent confounders. Most of the existing methods developed for this setting assume that the underlying model is linear Gaussian, including the Bayesian IDA (BIDA) method that we build upon in this work. By combining a Bayesian backdoor estimator with model averaging, we obtain a posterior over the intervention distributions of a cause-effect pair that can be expressed as a mixture over stochastic linear combinations of Dirichlet distributions. Although there is no closed-form expression for the posterior density, 
it is straightforward to produce Monte Carlo approximations of target quantities through direct sampling, and we also derive closed-form expressions for a few selected moments. To scale up the proposed procedure, we employ Markov Chain Monte Carlo (MCMC), which also enables us to use more efficient adjustment sets compared to the current exact BIDA. Finally, we use Jensen-Shannon divergence to define a novel causal effect based on a set of intervention distributions in the general categorical setting. We compare our method to the original IDA method and existing Bayesian approaches in numerical simulations and show that categorical BIDA performs favorably against the existing alternative methods in terms of producing point estimates and discovering strong effects.
\end{abstract}

\input{sections/intro}

\input{sections/causal_model}

\input{sections/posterior-params}

\input{sections/numerical_experiments}

\input{sections/discussion}

\section*{Funding}
This research was supported by the Research Council of Norway under the Centres of Excellence scheme (grant 332645: Integreat - The Norwegian centre for knowledge-driven machine learning).

\bibliographystyle{apalike}
\bibliography{library}

\clearpage
\section*{Supplements}
% Use letters for section numbering 
\renewcommand{\thesection}{\Alph{section}}
\newcommand*\rot{\rotatebox{90}}
\counterwithin{figure}{section}
\counterwithin{table}{section}
\setcounter{section}{0}

\input{\rootdir{sections/appendix-moments}}\clearpage

\input{\rootdir{sections/appendix-illustration-shape-of-posterior}}\clearpage   
\input{\rootdir{sections/appendix-sim-setup}}\clearpage       
\input{\rootdir{sections/appendix-sim-res-binary}}\clearpage  
\input{\rootdir{sections/appendix-sim-res-all}}\clearpage     

\end{document}

%% file: sections/intro.tex
\section{Introduction}
Causal inference is the process of determining how a change in one variable, through an intervention, affects the behaviour of other variables within a system. Understanding such causal relationships is the ultimate goal of many scientific studies and is essential for making informed decisions about the effect of future interventions. While the gold standard for inferring causal relationships is through controlled experiments, performing such experiments is often not feasible due to ethical, economical or simply practical reasons. Consequently, in many practical applications, one would like to infer causal relationships based on observational data, which has been obtained by passively observing the considered system. 

The problem of inferring causal relationships or effects from observational data has received a lot of attention during the last few decades \citep{Spirtes1993, pearlCausality2009, petersElementsCausalInference2017, hernanCausalInferenceWhat2023}. In particular, the graph-based approach of do-calculus has emerged as one of the main frameworks for causal reasoning in complex systems \citep{pearlCausality2009}. The key assumption in do-calculus is that one has access to a causal structure over the considered variables, typically represented by a directed acyclic graph (DAG).  The rules of do-calculus can then be applied to derive a non-parametric expression for a target intervention distribution that can be estimated from observational data, assuming that the intervention distribution in question is identifiable under the causal assumptions imposed by the DAG, that is, can be re-expressed in terms of standard conditional distributions.  
The most well-known example of such an expression is known as the backdoor adjustment formula \citep{pearlCausality2009}.

In a more challenging setting of causal inference, yet often most realistic, the causal structure is only partly or even completely unknown. In such a setting, the causal structure must then also be learned from the available data, a problem known as (causal) structure learning in the graphical model community \citep{Koller2009}. The vast number of possible structures makes structure learning a challenging problem already in the simpler probabilistic (non-causal) setting. Another challenge, specific to the causal setting, is that multiple DAGs can represent the same set of conditional independence statements, forming a so-called Markov equivalence class. A unique causal DAG can therefore not typically be learned from the independence patterns over the variables in the observed data. Exploiting these independence patterns, the best one can hope for is to infer the correct Markov equivalence class, and any subsequent causal inference then needs to be based on the entire class of equivalent DAGs.

%Even if the causal effects cannot be uniquely identified under the true observational distribution, such inference can still provide some insight and be highly useful. In particular, one might be able to infer what are likely or unlikely to be strong causal relationships in a complex system without much prior knowledge. This information can be used, for example, to guide future experiments designed to validate or refute the initial findings. This is especially relevant in the high-dimensional setting where intervening on every variable is often not be feasible. 

The ambition to learn causal effects from observational data under an unknown DAG can be approached through the idea of combining a structure learning procedure with a method for estimating causal effects under given a structure. 
As a first scalable method in this direction, \cite{Maathuis2009} introduced the IDA algorithm. The algorithm first learns an equivalence class from the data, and then, for each cause-effect pair, estimates the set of causal effects consistent with this class.
Considering the challenging nature of the structure learning task, the algorithm is prone to errors in the inferred structure. Typically, there will be multiple non-equivalent structures that fit the data almost equally well, yet carry very different causal assumptions. To better account for the structural uncertainty, IDA has been combined with various resampling strategies \citep{stekhovenCausalStabilityRanking2012, taruttisStatisticalApproachVirtual2015} and adapted to the Bayesian setting \citep{Pensar2020}. In this work, we build upon the latter approach, referred to as Bayesian IDA (BIDA), which was shown to outperform the original IDA method and the frequentist resampling variants in terms of accuracy.
Like the original IDA method, BIDA employs the backdoor formula to estimate causal effects locally in a given structure, albeit with a Bayesian estimation approach. In addition, Bayesian model averaging is used to account for the structure uncertainty.

A key assumption in the current BIDA algorithm is that the underlying model is linear Gaussian, which implies that the causal effects are linear and can be conveniently estimated using linear regression.
In this work, we assume a categorical model and extend BIDA to this setting.
More specifically, we assume that we have access to passively observed categorical data over a (potentially) high-dimensional causal system for which the underlying structure is completely unknown. Based on the available data, our ambition is to estimate the intervention distributions and causal effect for every pair of variables in the system. Importantly, we assume that the true causal structure can be represented by an unknown DAG and that there are no unobserved latent confounders.
Under these assumptions, we derive a Bayesian backdoor estimator for categorical intervention distributions, which is flexible with respect to the causal contrasts of interest. In particular, we focus on a novel causal effect for the general categorical setting, which we define as the Jensen-Shannon divergence (JSD) between the intervention distributions \citep{linDivergenceMeasuresBased1991, griffithsMeasuringCausalSpecificity2015}. 

The main challenge with the Bayesian approach is computational. Averaging over all possible structures becomes infeasible even for smaller-scale systems, since the size of the graph space grows super-exponentially with the number of variables. 
In its original version, the BIDA algorithm relies on a dynamical programming routine that computes exact posterior probabilities over parent sets, and it is applicable to systems with up to around 25 nodes. In order to scale up BIDA to larger systems, we make use of Markov Chain Monte Carlo (MCMC) for sampling DAGs from the graph posterior \citep{Koller2009, Kuipers2017, kuipersEfficientSamplingStructure2022, Viinikka2020}. In particular, we employ the partition-MCMC scheme \citep{kuipersEfficientSamplingStructure2022} to approximate the posterior over adjustment sets. This allows us not only to scale up the method, but also to consider various adjustment sets identifiable through graphical criteria, including the o-set \citep{henckelGraphicalCriteriaEfficient2022, Witte2020}. 

The local strategy for estimating causal effects distinguishes the BIDA method from other MCMC-based Bayesian approaches for learning causal effects under an unknown DAG, which have been proposed for both the Gaussian setting \citep{Viinikka2020, kuipersInterventionalBayesianGaussian2022a, Castelletti2021} and the binary setting \citep{moffaUsingDirectedAcyclic2017, castellettiJointStructureLearning2024}. These methods can be considered full Bayesian approaches, in a more conventional sense, in that they sample full models (that is, causal structures and associated model parameters) from the posterior, and then perform inference in the models to compute the target intervention distributions and causal effects. 
However, such inference can be computationally very demanding in the categorical setting and even intractable in larger networks, depending on the model structure. In comparison to this full Bayesian analysis, the BIDA posterior can be viewed as a scalable approximation that, under the proposed assumptions, will converge to the same target distribution.

The remainder of this paper is structured as follows. In Section \ref{sec::causal_model} we introduce the causal model and the assumptions required to make causal inference based on observational data. We also present the backdoor formula and different classes of backdoor adjustment sets. In Section \ref{sec::estimator} we present our BIDA procedure for Bayesian estimation of the target effects. In Section \ref{sec::num_exp} we report the results from an extensive simulation study. First, we compare BIDA with the existing full Bayesian approaches in the binary setting. We then move to the general categorical setting and apply BIDA on 9 benchmark networks, investigate the choice of adjustment set and compare BIDA against two non-Bayesian IDA variants. Finally, in Section \ref{sec::discussion} we conclude with a short discussion and provide some ideas for further development of the procedure.

%% file: sections/causal_model.tex
\section{Causal Bayesian networks and intervention distributions\label{sec::causal_model}}
\subsection{The causal model}
 Let $G = (V, E)$ be a directed acyclic graph (DAG) consisting of a set of nodes $V$ and a set of directed edges $E\subset V\times V$, such that there are no directed cycles in $G$. If there is an edge from $i$ to $j$ in $G$, $i$ is said to be a parent of $j$. We let $G_i$ denote the set of parent nodes of $i$ in $G$. If there is a directed path from $i$ to $j$, we say that $i$ is an ancestor of $j$ and that $j$ is descendant of $i$. 
 
 In probabilistic DAG models, known as Bayesian networks, the nodes represent random variables and the edges represent probabilistic dependencies between the variables. We let $V$ denote the index set $\{1, ..., n\}$ and associate with each node $i \in V$ a variable $X_i$. By the local Markov condition  \citep[Ch 3]{Koller2009}, the joint distribution of a Bayesian network factorizes over $G$ according to 
\begin{align}
    P(x_1, ..., x_n) &=\prod_{i=1}^nP(x_i\vert x_{G_i}), \label{eq::markov}
\end{align}
where $x_{G_i}$ is a particular instance of the set of variables associated with the parent set $G_i$.  
Using the graphical criterion $d$-separation, one can directly read off $G$ all conditional independencies in $P$ that are imposed by $G$. If there are no additional independencies in $P$, $P$ is said to be faithful to $G$.

When given a causal interpretation, an edge $i\rightarrow j$ in the DAG represents a direct causal relationship from $X_i$ to $X_j$ with respect to the considered variables. The DAG does then not simply describe the independence properties of the model, as captured by Equation \eqref{eq::markov}, but it also carries assumptions about the world and the causal mechanisms that ultimately produce these independencies. Importantly, a causal DAG is assumed complete in the sense that it includes all common causes of any pair of variables in the system, and that all causal relations between the variables are correctly specified by its edges  \citep{Spirtes1993}. For inferential purposes, we will make the additional assumption of causal sufficiency, implying that all common causes are not only included in the model, but also observed in the data (i.e. no unobserved confounders). Moreover, we will assume that the distribution $P$ is faithful to $G$, which is to say that all independencies in $P$ are due to the (lack of) causal relations in $G$, and that the distribution is positive, that is, $P(x) > 0 $ for every joint configuration $x$.  

Under the given assumptions, the model can be used to represent and predict the effect of interventions on one or several variables. We use the do-operator, and write $do(X_i = x_i')$, or simply $do(x_i')$, to denote that $X_i$ is set to the value $x_i'$ by an external intervention that, importantly, leaves all other variables in the model undisturbed. 
\noindent The distribution under such an intervention is given by the truncated factorization formula  \citep[70-74]{pearlCausality2009}:
\begin{align} \label{eq::trunc}
	P(x_1, ..., x_n\vert do(x_i')) =\begin{cases}
	  \prod_{j\neq i}P(x_j\vert x_{G_j})  &\text{if $x_i = x_i'$,} \\
	0 \quad &\text{otherwise}, \\
	\end{cases}
\end{align}
where the product on the right-hand side is made up of the conditional probabilities from the pre-intervention distribution defined by our causal model in \eqref{eq::markov}. Thus, under the given assumptions, the intervention distribution can be estimated consistently from observational data. Furthermore, all relevant intervention distributions can be computed by marginalizing out irrelevant variables from the appropriate truncated factorization.

In this work we are considering the task of estimating the causal (intervention) effects between every cause-effect pair in the considered system, that is, for each pair $(i, j) \in V\times V$. The considered causal effect is defined in terms of intervention distributions and measures how strongly interventions on $X_i$ affect the marginal distribution of $X_j$. In other words, our target intervention distributions are the marginal distributions of the form $P(X_j|do(x_i'))$ under each possible intervention $x_i'$.
As described above, the intervention distributions can be computed from an estimate of the complete model through marginalization of the truncated factorization \eqref{eq::trunc}. An alternative and computationally more efficient approach is to estimate the target distributions more directly through the backdoor adjustment formula, which in the categorical setting can be expressed as: 
\begin{align}\label{eq::backdoor}
	P({x}_j\vert do(x_i')) =\begin{cases} 
        P(x_j), &\text{if $ j \in G_{ij}$},\\
        \sum_{x_{G_{ij}}} P({x}_j\vert  x_i', x_{G_{ij}})P(x_{G_{ij}}), &\text{otherwise,}
	\end{cases}
\end{align}
where $G_{ij}$ is a valid backdoor adjustment set with respect to the cause-effect pair $(i,j)$ in $G$.
Valid adjustment sets can be identified directly from the graph using certain graphical criteria  \citep{pearlCausality2009, perkovicCompleteGraphicalCharacterization}. We delve into these criteria and characterize four classes of backdoor adjustment sets in the next section. In the special case where $G_{ij}$ includes node $j$, interventions on $X_i$ has no effect on $X_j$.
Otherwise, averaging over the adjustment set through the above formula removes the confounding effect of any potential common causes of $X_i$ and $X_j$.

\subsection{Backdoor adjustment sets}\label{sec::backdoor_sets}
A backdoor adjustment set is defined as a set of nodes that can be used to adjust for confounding through the backdoor formula \eqref{eq::backdoor}. 
Such sets are characterized by certain graphical criteria, that can be derived through do-calculus.  
Following Pearl's (\citeyear{pearlCausality2009}) original definition, a set $G_{ij}$ is a valid adjustment set with respect to a cause-effect pair $(i,j)$ in a causal DAG $G$ if:  (i) $G_{ij}$ contains no descendants of node $i$; and (ii) $G_{ij}$ d-separates node $i$ and $j$ in the backdoor graph, which is obtained from $G$ by removing all edges that go out from $i$.
We refer to such sets as backdoor adjustment sets or simply adjustment sets.
 
In general, there will be more than one possible adjustment set. While any valid adjustment set produces a consistent estimator, different sets will result in more or less efficient estimators from a statistical point of view. In the high-dimensional setting, the computational efficiency of the adjustment set is also a relevant factor to consider. Here, we consider four classes of backdoor adjustment sets: 
\begin{itemize}%[noitemsep]
    \item the parent set of the cause node; 
    \item the o-set  \citep{henckelGraphicalCriteriaEfficient2022}; 
    \item the minimal parent set;
    \item the minimal o-set.
\end{itemize}
In a DAG, the o-set is defined as the parents of all nodes lying on directed paths between the cause node $i$ and the effect $j$ (so-called ``causal paths''), excluding parents that are descendants of $i$ (so-called ``forbidden nodes'').
If there are no causal paths between $i$ and $j$, the o-set is not defined. We then set $G_{ij}=\{j\}$, such that the backdoor adjustment formula in \eqref{eq::backdoor} returns the marginal pre-intervention distribution of $X_j$.
The term ``minimal'' refers to a reduced set from which no node can be removed without rendering the set invalid for adjustment. Since neither the parent set nor the o-set is in general minimal, we also consider minimal versions of them, as defined formally below.

\begin{definition}
Let $G = (V, E)$ be a causal DAG, and let $G^{\underline{i}}$ be the backdoor graph from which all edges going out from $i$ have been removed. For every pair $(i, j) \in V\times V$ for which there exists a causal (directed) path from $i$ to $j$ in $G$, let $G_{ij}$ denote the full parent set (o-set). The minimal parent set (o-set) is then defined as the subset $G_{ij}^\text{min} \subseteq G_{ij}$ such that, for every node $v \in G_{ij}^\text{min}$, the set $G_{ij}^\text{min} \setminus \{v\}$ does not d-separate $i$ and $j$ in the backdoor graph $G^{\underline{i}}$. For pairs $(i, j)$ not connected by a causal path, we set $G^\text{min}_{ij} =\{j\}$, encoding that interventions on $X_i$ has no effect on $X_j$. 
\end{definition}
\noindent When there exists a causal path from $i$ to $j$, the minimal parent set and the minimal o-set are examples of minimal separators in the backdoor graph, and both are unique. They then coincide if there are no backdoor paths connecting $i$ and $j$ (no confounding) and the empty set is the unique minimal separator, or if there are no non-descendants of $i$ on any backdoor path, except from  $i$'s parents.  Efficient algorithms for computing minimal separators can be found, for example, in \citet{zanderFindingMinimalDseparators2020}. 

The four classes of backdoor adjustment sets listed above have different theoretical properties with respect to statistical and computational efficiency.
The parent set $G_i$ of a cause node $i$ is a valid adjustment set for all its potential effects $j$ and can be identified locally simply by considering the neighbours of a node in the DAG. 
In contrast, the o-set and the minimal adjustment sets are specific to every cause-effect pair $(i,j)$ and can not be determined locally in the DAG, as a global search for a directed path from node $i$ to $j$ is required. 
From a statistical perspective, on the other hand, the parent set is not the most efficient choice of adjustment set. In estimating the mean-difference between intervention distributions, the o-set has been shown to have the lowest asymptotic variance among all valid adjustment sets, assuming the true DAG is known  \citep{henckelGraphicalCriteriaEfficient2022, rotnitzkyEfficientAdjustmentSets2020}. Its minimal variance property is, however, not guaranteed in finite samples \citep{kuipersVarianceCausalEffect2022} or under model uncertainty. Simulations studies suggests that adjusting for covariates strongly associated with the outcome rather than the cause improves estimation accuracy in various causal inference settings  \citep{witteCovariateSelectionStrategies2019, Witte2020}. Intuitively, the o-set contains nodes as close to $j$ as possible, and should be preferable as such, assuming the DAG is correctly specified. 
Yet, under model uncertainty, the parent sets are possibly more robust to partial misspecification of the causal structure. If a given structure correctly specifies the true parents of $i$, but wrongly states that there is no causal path from $i$ to $j$ due to a single misoriented edge, using parents for adjustment will still yield a consistent estimator. 

The inclusion of minimal adjustment sets is motivated by the intuition that distributions over smaller sets are easier to estimate, especially in the categorical setting where data fragmentation becomes an issue. In high-dimensional settings, there is also a computational argument for limiting the size of adjustment sets by removing redundant covariates from the backdoor formula.

\subsection{Quantifying causal effects between categorical variables \label{sec::JSD}}
Defining causal effects in terms of interventions, we say that there is a causal effect of $X_i$ on $X_j$ if interventions on $X_i$ changes the distribution over $X_j$. To quantify the effect we want to measure by how much interventions on $X_i$ change the distribution of $X_j$ and summarize these changes by a single quantity $\tau_{ij}$. Assuming a binary (or continuous) $X_j$, the causal effect is often defined as the expected change in $X_j$ under two distinct intervention levels $x_i'$ and $x_i''$ \citep{pearlCausality2009}: 
\begin{align}
E[X_j| do(X_i = 1)] - E[X_j| do(X_i = 0)]. \label{eq::ATE}
\end{align}
When $X_i$ represents some binary treatment, this difference is commonly referred to as the average treatment effect (ATE). In the general discrete case, however, we need an alternative measure that can account for more than two outcomes on both the cause and effect variable.

Here, we define the causal effect in terms of the distance between intervention distributions.
Specifically, we define $\tau_{ij}$ as the Jensen-Shannon divergence (JSD) for the set of intervention distributions over $X_j$ induced by each possible intervention on $X_i$:
\begin{align}
    \tau_{ij}  & = \sum_{x_i}\sum_{x_j}\omega_{x_i}P(x_j|do(x_i)) \log\frac{P(x_j|do(x_i)) }{\sum_{x_i}\omega_{x_i}P(x_j|do(x_i))},  \label{eq::JSD}
\end{align}
where $\omega_{x_i}$ denotes a weight assigned to the associated intervention distribution. The JSD is the (weighted) average Kullback-Leibler divergence between each intervention distribution and the (weighted) average of these distributions. It is as such non-negative and zero only if the contrasted distributions are equal, corresponding to the case where interventions on $X_i$ do not affect the distribution of $X_j$. The JSD is also symmetric, a desired property in this setting, and bounded from above, as:
\begin{align*}
0 \leq \tau_{ij}  \leq -\sum_{x_i} \omega_{x_i} \log(\omega_{x_i}).
\end{align*}
The upper bound is reached when distinct outcomes of $X_j$ occur with certainty under each possible intervention. The set of intervention distributions then forms a bijective mapping from $X_i$ to $X_j$, and could not be more different. 

To the best of our knowledge, the JSD has not previously been used to quantify causal effects in the causal discovery literature. However, similar information theoretic tools are used to formalize concepts of causal influence (see for example \citealt{janzingQuantifyingCausalInfluences2013, griffithsMeasuringCausalSpecificity2015, wieczorekInformationTheoreticCausal2019}).
In particular, \citet{griffithsMeasuringCausalSpecificity2015} suggest $\tau_{ij}$ as a measure of \textit{causal specificity}, a concept regarding to which extent a cause controls an outcome. They motivate it as the mutual information between a random intervention and an outcome as follows. Suppose an intervention on $X_i$ sets its value randomly, such that $X_i = x_i$ with probability $w_{x_i}$. This random intervention produces a mixture distribution for $X_j$, namely the average intervention distribution $\sum_{x_i}\omega_{x_i}P(x_j|do(x_i))$.  Now, if interventions on $X_i$ do affect $X_j$, the uncertainty about $X_j$ is reduced if one is informed about the exact intervention level $x_i$. The expected information gain is then measured by the mutual information between the random intervention on $X_i$ and the outcome $X_j$, which can be shown to be equal to $\tau_{ij}$ as defined in \eqref{eq::JSD}. 

The above interpretation of $\tau_{ij}$ also provides a way of reasoning about the weights in \eqref{eq::JSD}, as these describe under which random intervention scheme we would expect to observe $\tau_{ij}$. If non-uniform, some interventions are assumed more likely than others. 
We will use uniform weights, as our goal is to predict the effects of interventions we would observe in a future experiment, where one would be in perfect control of the intervention variable and where we consider every intervention equally important or likely. 
The choice of weights does, however, also have statistical consequences. Intervention distributions associated with levels of $X_i$ that are never or rarely observed are more difficult to estimate.
A more pragmatic approach would therefore be to assign each distribution a weight proportional to the observed sample counts of $X_i$, as proposed by \citet{janzingQuantifyingCausalInfluences2013}. The future interventions on $X_i$ should then be designed to mimic its behaviour in a passively observed system free from interventions. While statistically more convenient, we argue that such an approach is difficult to motivate in our context. 

\onlyifstandalone{\bibliographystyle{apalike}\bibliography{\rootdir{library}}}

%% file: sections/posterior-params.tex
\section{Bayesian estimation of intervention distributions and causal effects\label{sec::estimator}} 
In this section, we extend the Bayesian IDA (BIDA, \citet{Pensar2020}) to the categorical setting and combine it with an approximate MCMC based model averaging approach that allows us both to scale up and to use more efficient adjustment sets compared to the original method. As such, we present a Bayesian estimator of intervention distributions and the associated causal effects for each cause-effect pair in a system, under the assumption that we are given a set of observational categorical data $D$ sampled from an unknown causal Bayesian network.

Specifically, let $X_1, \ldots, X_n$ be a set of categorical variables, each of which can take one of $r_i\geq 2$ possible values: $x_i^1, \ldots, x_i^{r_i}$. Then, for every $(i, j) \in V\times V$, our goal is to estimate the associated marginal intervention probabilities:
\begin{equation*}
\pi_{x_j|x_i} = P(x_j|do(x_i)), \quad x_i = x_i^1, \ldots, x_i^{r_i},\; x_{j} = x_{j}^1, \ldots, x_{j}^{r_{j}},
\end{equation*}
where $\sum_{x_j} \pi_{x_j|x_i} = 1$. We denote a specific intervention distribution by
\begin{equation*}
\pi_{X_j|x_i} = (\pi_{x^1_j|x_i}, \ldots , \pi_{x^{r_j}_j|x_i}),
\end{equation*} 
and the full collection of intervention distributions for a given cause-effect pair:
\begin{equation*}
\pi_{ij} = \big\{\pi_{X_j|x_i^k}\big\}_{k = 1}^{r_i},
\end{equation*} 
which we refer to as the intervention probability table (IPT) for pair $(i,j)$. %Finally, we denote the complete collection IPTs by 
%\begin{equation*}
%\pi = \big\{\pi_{ij}\big\}_{(i,j)\in V \times V}.
%\end{equation*} 

Using the above notation, the complete set of causal parameters we seek to estimate is thus the $n\times n$ IPT matrix $\pi = \{\pi_{ij}\}$ and then, ultimately, the corresponding causal effect matrix $\tau = \{\tau_{ij}\}$ which is a function of the former. To this end, we start by presenting our Bayesian estimator of the IPT matrix in the simpler setting where the causal structure, $G$, is assumed known. Then, we extend our estimator to the setting where $G$ is unknown. Finally, we present our Bayesian estimator for the causal effect matrix.

\subsection{Posterior of intervention probabilities under a given structure \label{sec::posterior-params-known-dag}}
Under the given assumptions and a known (and correct) causal DAG $G$, valid adjustment sets can be identified and the backdoor formula \eqref{eq::backdoor} applied to consistently estimate intervention distributions from observational data. In particular, rather than fitting a full causal model and performing inference within that model, one can target the marginal intervention distributions, $\pi_{ij},$ more directly by estimating the (conditional) distributions involved in the backdoor formula of each cause-effect pair. Following this more local approach, we construct a Bayesian estimator for the posterior distributions over $\pi_{ij}$ given $G$ as follows. 

For each pair $(i, j)$, suppose that $G_{ij}$ is the adjustment set of a predefined class in $G$. 
First, consider the case where $j \notin G_{ij}$, that is, the case where $X_i$ is possibly a cause of $X_j$ and the intervention distribution is not reduced to the marginal distribution. 
Let $\theta_{ij}$ denote the parameters of the observational distributions involved in the backdoor formula \eqref{eq::backdoor} for cause-effect pair $(i,j)$ under graph $G$,\footnote{To avoid unnecessarily cluttered notation, we do not explicitly include the dependence on $G$ in the specification of the parameter vectors contained by $\theta_{ij}$.} that is, the conditional distributions:
\begin{equation*}
    \theta_{X_j|x_i, x_{G_{ij}}} = (\theta_{x_j^1| x_i, x_{G_{ij}}}, \ldots, \theta_{x_j^{r_j}|x_i, x_{G_{ij}}}), \quad x_i = x_i^1, \ldots, x_i^{r_i},\; x_{G_{ij}} = x_{G_{ij}}^1, \ldots, x_{G_{ij}}^{r_{G_{ij}}},
\end{equation*}
as well as the marginal distribution over the adjustment set:
\begin{equation*}
    \theta_{X_{G_{ij}}} = (\theta_{x^1_{G_{ij}}}, \ldots, \theta_{x^{r_{Gij}}_{G_{ij}}}).
\end{equation*}
Now, we specify a joint prior distribution directly over the backdoor parameter vectors that is identical for each $G$ with adjustment set $G_{ij}$ and factorizes according to: $$p(\theta_{ij} | G)= p(\theta_{X_{G_{ij}}}) \prod_{x_i,x_{G_{ij}}} p(\theta_{X_j\mid x_i, x_{G_{ij}}}).$$ 
Moreover, we assume a Dirichlet prior for each parameter vector, 
with hyperparameters: 
\begin{align*}
    \alpha_{X_j|x_i, x_{G_{ij}}} &= (\alpha_{x_j^1\vert x_i, x_{G_{ij}}}, ..., \alpha_{x_j^{r_j}\vert x_i, x_{G_{ij}}}), \\
    \alpha_{X_{G_{ij}}} &= (\alpha_{x^1_{G_{ij}}}\;, ..., \alpha_{x^{r_{Gij}}_{G_{ij}}}),
\end{align*}
respectively, such that $\alpha_{x_{G_{ij}}} = \sum_{x_i}\sum_{x_j} \alpha_{x_j\vert x_i, x_{G_{ij}}}$. Specifically, we set the hyperparameters to
\begin{align*}
    \alpha_{x_j\vert x_i, x_{G_{ij}}} = \alpha_0 P_0(x_j, x_i, x_{G_{ij}}) = \frac{\alpha_0}{r_jr_ir_{G_{ij}}},
\end{align*}
given an equivalent sample size $\alpha_0$ and a prior distribution $P_0$, here assumed uniform. 
Note that the only structural information reflected in this prior distribution is that $G_{ij}$ is a valid adjustment set. The prior does not encode any independence restrictions between the involved variables, even if such restrictions would be implied by $G$. In fact, it coincides with the beliefs stated in a BDeu-prior \citep{heckermanLearningBayesianNetworks1995} for a complete DAG over nodes $(i, j, G_{ij})$.\footnote{A complete DAG does not encode any conditional independencies. The same restriction-free prior can, in terms of the full network, be specified as a Dirichlet prior over the probabilities for each joint outcome over all the $n$ variables: 
$$p(\theta_{X_1, \ldots, X_n}) = \text{Dir}(\alpha_{X_1, \ldots, X_n}), $$
The priors over $\theta_{X_j|x_i, x_{G_{ij}}}$ and $\theta_{X_{G_{ij}}}$ would then follow from the aggregation properties of the Dirichlet distribution, for all combinations of $(i, j, G_{ij})$.}
%To emphasize the independence of $G$, we write $\tilde\theta$ %for parameters that do not obey any independence restrictions.}

Under the above prior, which is conjugate for the categorical likelihood, the posterior distributions of the considered parameter vectors are conveniently given by:
\begin{equation}\label{eq::post_params}
\begin{aligned}
    p( \theta_{X_j|x_i, x_{G_{ij}}}|D) &= \text{Dir}(\alpha_{X_j|x_i, x_{G_{ij}}} + N_{X_j|x_i, x_{G_{ij}}}), \\
    p( \theta_{X_{G_{ij}}}| D) &=\text{Dir}(\alpha_{X_{G_{ij}}} + N_{X_{G_{ij}}}), 
\end{aligned}
\end{equation}
where $N_{X_j|x_i, x_{G_{ij}}}$ and $N_{X_{G_{ij}}}$ are vectors of observed sample counts. More precisely, if $N_{x_j\vert x_i,x_{G_{ij}}}$ equals the number of samples in $D$ for which $X_j = x_i,\; X_i = x_i$ and $X_{G_{ij}} = x_{G_{ij}}$, then
\begin{align*}
    N_{X_j| x_i, x_{G_{ij}}} &= 
    (N_{x_j^1\vert x_i,x_{G_{ij}}}, \ldots, N_{x_j^{r_j}\vert x_i, x_{G_{ij}}}), \\
    N_{X_{G_{ij}}} &= (N_{x^1_{G_{ij}}}, \ldots, N_{x^{r_{Gij}}_{G_{ij}}}).
\end{align*}

For a given intervention level $x_i$, the associated intervention distribution can be expressed through the backdoor formula \eqref{eq::backdoor} using our observational parameters:
$$\pi_{X_j|x_i} = \sum_{x_{G_{ij}}}\theta_{X_j|x_i, x_{G_{ij}}}\theta_{x_{G_{ij}}},$$
which is then a stochastic linear combination of Dirichlet vectors, unless the adjustment set $G_{ij}$ is empty. The exact distribution of such linear combinations are, in general, not tractable.\footnotemark
%Moreover, parameters $\tilde\pi_{X_j|x_i}$ and $\tilde\pi_{X_j|x_i'}$ for two distinct intervention levels $x_i$ and $x_i'$ are dependent through the dependence on  $\tilde\theta_{X_{G_{ij}}}$. 
Analytical expressions for the posterior mean and covariance between each pair of elements in $\pi_{ij}$ can, however, be derived and are given at the end of this section.
When the adjustment set is empty, the intervention distribution $\pi_{X_j|x_i}$ equals the observational counterpart $\theta_{X_j|x_i}$. The IPT $\pi_{ij}$ is then a set of independent Dirichlet distributed vectors. 
\footnotetext{There are settings where a stochastic linear combination of Dirichlet vectors reduces to a Dirichlet vector. For example, if the updated hyperparameters had satisfied
$$\alpha_{x_{G_{ij}}} + N_{x_{G_{ij}}} = \sum_{x_j} (\alpha_{x_j\vert x_i, x_{G_{ij}}} + N_{x_j\vert x_i, x_{G_{ij}}}),$$
the distribution of $\pi_{X_j|x_i}$ can be shown to be equal to a Dirichlet vector \citep{Homei2021}. However, for a non-empty adjustment set $G_{ij}$, this restriction is not satisfied under the considered prior.}

For the case where $j \in G_{ij}$, each row in $\pi_{ij}$ equals the observational probabilities $\theta_{X_j}$. Analogously to the above, we assume a conjugate Dirichlet prior for $\theta_{X_j}$ with hyperparameters $\alpha_{x_j} = \alpha_0/r_j$, resulting in 
\begin{align}\label{eq::post_params_xj}
	p(\theta_{X_j}|D) &= \text{Dir}(\alpha_{X_j} + N_{X_j}). 
\end{align}

The target conditional posterior distribution $p(\pi_{ij}|G_{ij}, D)$ is then defined by:
\begin{align} \label{eq::backdoor_param}
\pi_{X_j|x_i} = 
\begin{cases} 
\theta_{X_j}&\text{if $j \in G_{ij}$,} \\
 \sum_{x_{G_{ij}}}\theta_{X_j|x_i, x_{G_{ij}}}\theta_{x_{G_{ij}}} &\text{otherwise,}
\end{cases}    
\end{align}
where the observational parameters are distributed according to \eqref{eq::post_params} and \eqref{eq::post_params_xj}. 
Note that it is straightforward to sample directly from the distribution and compute Monte Carlo approximations of relevant target quantities, even when there is no tractable closed-form expression for the exact distribution.

While the direct estimation approach outlined above will yield consistent estimates, it does ignore causal information in the DAG that we, for now, have assumed known.
In particular, even if the DAG states that $X_i$ is not a cause of $X_j$, the estimated intervention distributions $\pi_{X_j|x_i}$ may differ across intervention levels $x_i$ (implying a causal effect not exactly zero), because the independencies encoded in the DAG do not hold exactly in finite samples. The notable exception is when $j$ is included in the adjustment set $G_{ij}$, indicating that $X_i$ has no effect on $X_j$ and that the intervention distributions reduce to the marginal distribution according to \eqref{eq::backdoor_param}.
Hence, we can easily account for this information by setting the adjustment set $G_{ij} = j$ when there is no causal path from $i$ to $j$ in $G$.
Among the adjustment set we consider in this work, only the parent set is defined locally in $G$ and does not imply such conditioning on causal paths (see Section \ref{sec::backdoor_sets}). The three alternative classes of adjustment set will set effects to zero exactly given a structure where no causal path exists.

\subsubsection{Moments of the posterior \label{sec::moments}}
Even if there is no general tractable closed-form expression for the IPT posterior, $p(\pi_{ij}|G, D)$, expressions for certain moments can be derived analytically. In particular, we give here the mean $E[\pi_{x_i|x_j}]$ of every element and the covariance $Cov(\pi_{x_j|x_i}, \pi_{x_j'|x_i'})$ between any pair of elements in $\pi_{ij}$.
To simplify the notation, we specify these moments in terms of an intervention variable $X = X_i$, an effect variable $Y = X_j$, and a non-empty set of valid adjustment variable(s) $Z = X_{G_{ij}}$ and write $\pi_{y|x} = P(y|do(x))$ for the associated intervention probability. 

\sloppy
\begin{theorem}\label{prop::moments}
Assume that the random vectors $\theta_{Y|x^1, z^1}, \ldots, \theta_{Y|x^{r_x}, z^{r_z}}$ and $\theta_Z$ are mutually independent and that each vector is Dirichlet distributed:
\begin{equation*}
\begin{aligned}
    p(\theta_{Y|x, z}) &= \text{Dir}(\alpha_{y^1\vert  x,z}, \ldots, \alpha_{y^{r_y}\vert  x,z}), \ \ \text{ for each } (x,z), \\
    p(\theta_{Z}) &=\text{Dir}(\alpha_{z^{1}}, \ldots, \alpha_{z^{r_z}}). 
\end{aligned}
\end{equation*}
such that $\alpha_{z} = \sum_x\sum_y \alpha_{y|x,z}$ for all $z$. Then, the random variables that result from the linear combinations: $$\pi_{Y|x} = \sum_z \theta_{Y|x, z}\theta_z, \ \ \text{ for each } x,$$ have the following moments:
\begin{itemize}
\item[(i)] The mean of each element $\pi_{y|x}$ is given by:
$$
E[\pi_{y|x}] =\sum_z  \tilde\alpha_{y|x, z}\tilde\alpha_z,
$$ 
where $\tilde\alpha_{y|x,z} = \alpha_{y|x,z}/\sum_{y'}\alpha_{y'\vert  x,z}$ and $\tilde\alpha_{z} = \alpha_{z}/\alpha$.
\item[(ii)] The covariance between a pair of elements $(\pi_{y|x}, \pi_{y'|x})$ is given by:
\begin{align*}
 Cov& (\pi_{y|x}, \pi_{y'|x}) = \\ 
&\frac{1}{\alpha+1}\left(\sum_z\frac{\tilde\alpha_{y|x, z}\tilde\alpha_z}{\alpha_{x, z}+1}\left(\delta_{yy'}(1+\alpha_z)+\tilde\alpha_{y'|x, 
z}(\alpha_{x, z}-\alpha_{z})\right) - E[\pi_{y|x}]E[\pi_{y'|x}]\right),
\end{align*}
where $\alpha = \sum_z\alpha_z$, $\alpha_{x,z} = \sum_y \alpha_{y\vert  x, z}$ and $\delta_{yy'} = 1$ if $y = y'$, and $\delta_{yy'} = 0$ otherwise.
\item[(iii)] The covariance between a pair of elements $(\pi_{y|x}, \pi_{y'|x'})$, for which $x \neq x'$, is given by:
\begin{align*}
Cov(\pi_{y|x}, \pi_{y'|x'}) 
&= 
 \frac{1}{\alpha+1}\left(\sum_z\tilde\alpha_{y|x,z}\tilde\alpha_{y'|x',z}\tilde\alpha_z  -E[\pi_{y|x}]E[\pi_{y'|x'}] \right).
\end{align*}
\end{itemize}
\end{theorem}

\begin{proof}
A full derivation of these moments is included in Appendix A in the supplement \citep{supplement}.
\end{proof}

%%%

\subsection{Posterior of intervention probabilities under an unknown structure \label{section::posterior-of-intervention-probabilities}}

The estimator outlined in the previous section can readily be used to estimate IPTs in any given DAG, under the given assumptions. However, our main interest is the more challenging setting where the underlying DAG is unknown, and hence has to be inferred from the available data, a task known as structure learning. 
The Bayesian approach to the structure learning problem targets the posterior distribution over the DAG structures, which describes the probability that the independencies implied by each structure are true given the observed data \citep{Heckerman2006}. Inference about a target parameter can then be made by averaging over all possible DAGs. 

In our context, the target parameter is the IPT, $\pi_{ij}$, and the posterior can be expressed as
\begin{align*}
    p(\pi_{ij}|\, D) = \sum_{G} p(\pi_{ij}|\, D,G)P(G|\, D),
\end{align*}
\sloppy
where $p(\pi_{ij}|\, D, G)$ is the IPT posterior for a given DAG $G$ and $P(G|\, D)$ is the posterior probability of that DAG. By Bayes rule, we have that $$P(G|\, D) \propto P(G) P(D|G)$$ where $P(G)$ is the prior probability of $G$ and $$P(D|G) = \int_{\theta^G} P(D|G, \theta^G)p(\theta^G|G)d\theta^G$$ is the marginal likelihood of the data given $G$, obtained by integrating over the model parameters, \begin{equation*}
\theta^G = \big\{ \theta_{X_i | X_{G_i}} \big\}_{i=1}^n,
\end{equation*} 
that is, the conditional distributions in the Bayesian network factorization \eqref{eq::markov}.
In terms of the parameter prior, $p(\theta^G|G)$, we assume a standard BDeu-prior \citep{heckermanLearningBayesianNetworks1995}, in which the parameter vectors $\theta_{X_i|x_{G_i}}$ for all nodes $i$ and parent configurations $x_{G_i}$ are assumed mutually independent, and each vector is assigned a Dirichlet prior with hyperparameters: 
\begin{align*}
    \alpha_{x_i| x_{G_i}} = \frac{\alpha_0}{r_ir_{G_i}}.
\end{align*}
Under the assumed prior, which is conjugate for the categorical likelihood, the marginal likelihood can be conveniently expressed in closed form as a product over the nodes and parent configurations:
\begin{equation*}
P(D|G) = \prod_{i = 1}^n\prod_{x_{G_i}} \frac{\Gamma(\alpha_{x_{G_i}})}{\Gamma(\alpha_{{x_{G_i}}}+N_{x_{G_i}})} \prod_{x_i}\frac{\Gamma(\alpha_{{x_i\vert x_{G_i}}}+N_{x_i\vert  x_{G_i}})}{\Gamma(\alpha_{{x_i\vert  x_{G_i}}})}. 
\end{equation*}
Importantly, the marginal likelihood under a BDeu prior satisfies score equivalence, meaning that any two distinct DAGs that belong to the same Markov equivalence class obtain the same marginal likelihood. This is particularly important in causal structure learning, where Markov equivalent DAGs can have very different causal interpretations, although being identical as representations of dependence structures. Finally, by assuming a score equivalent DAG prior that factorizes over the nodes, we obtain a posterior probability that satisfies score equivalence and factorizes over the nodes:
\begin{align*}
    P(G|\, D) \propto P(G) P(D|G) = \prod_{i = 1}^n S_i(G_i, D),
\end{align*}
such that each node-specific score or weight $S_i( G_i|D)$ can be computed independently for each node $i$ and parent set $G_i$. 

An important thing to note at this point is that the conditional distributions included in the set of backdoor parameters, $\theta_{ij}$, as well as the IPT $\pi_{ij}$, are in principle functions of the model parameters, $\theta^G$.  In a standard Bayesian approach for estimating $\pi_{ij}$, as proposed by \cite{moffaUsingDirectedAcyclic2017} and \cite{castellettiJointStructureLearning2024}, one would consider a single posterior over the full model $(G, \theta^G)$ that defines the target posterior $p(\pi_{ij}|G, D)$ via the truncated factorization \eqref{eq::trunc} or, equivalently, the backdoor adjustment formula \eqref{eq::backdoor}. However, in order to compute $\theta_{ij}$ from $\theta^G$, one needs to perform inference in the model, a task that is computationally demanding or even intractable, depending on the structure of the given model, limiting the scalability of the resulting inference method.

To bypass the need to perform inference in the models, we follow the Bayesian IDA approach of \cite{Pensar2020}. 
Specifically, we treat $\theta_{ij}$ and $\theta^G$ as independent parameter sets, decoupling the estimation of the DAG posterior, $P(G|D)$, and IPT posteriors under a given DAG, $p(\pi_{ij}|\, D,G)$. In contrast to a standard full Bayesian approach, we only make use of the structural information carried by a causal model to select adjustment sets, and then we ``re-estimate'' the posterior of $\theta_{ij}$ (and thus the posterior of $\pi_{ij}$) under a prior that is free from any independence restrictions. 
The resulting BIDA posterior $p(\pi_{ij}|\, D)$ can then be expressed as mixture distribution over adjustment sets:
\begin{align}
   p(\pi_{ij}|\, D) = \sum_{G_{ij}} p(\pi_{ij}|\, D,G_{ij})P(G_{ij}|\, D)\label{eq::posterior_pi},
\end{align}
where $P(G_{ij}|\, D)=\sum_{G: G_{ij}}P(G|D)$ is the posterior probability of $G_{ij}$ with respect to a predefined class of adjustment sets. 
Exact evaluation of these mixture components are, nevertheless, also demanding. We will rely on Markov Chain Monte Carlo (MCMC) methods to approximate this posterior, described in detail below. 

The BIDA posterior can be viewed as a scalable approximation of the posterior obtained under a conventional full Bayesian approach, in the sense that the BDeu prior used for computing the DAG posterior does not, in general, coincide with the parameter prior used for estimating the IPTs in the same DAG (see Section \ref{sec::posterior-params-known-dag}).
From an asymptotic perspective and under the given assumptions (including score equivalence), we note that the DAG posterior in theory will converge to a distribution where the probability mass is shared equally among the members in the true equivalence class. Hence, the weight of each adjustment-set-specific mixture component in \eqref{eq::posterior_pi}, $P(G_{ij}|D)$, will tend to the fraction of DAGs in the true equivalence class that agree on that particular adjustment set. If the true equivalence class include adjustment sets that lead to different estimates for $\pi_{ij}$, the posterior $p(\pi_{ij}|D)$ will remain multimodal even in the large sample limit, reflecting the inherent non-identifiability issue caused by Markov equivalence.  On the other hand, if all adjustment sets in the equivalence class result in the same effect under the true distribution, the posterior distribution of $\pi_{ij}$ will converge to a point mass at the true value, given that the posterior over the observational parameters in $\theta_{ij}$ converges to point masses at their true values.

Before we move on to describe how we approximate the posterior over adjustment sets, $P(G_{ij}|D)$, we note that the mean of the BIDA posterior in \eqref{eq::posterior_pi} can be readily computed applying Theorem \ref{prop::moments}.

\theoremstyle{plain}
\newtheorem{corollary}{Corollary  \label{corollary-posterior-mean}}[theorem]
\sloppy
\begin{corollary}
Assume that $\pi_{ij}$ given $D$ and $G_{ij}$ is distributed as in Equations \eqref{eq::post_params}-\eqref{eq::backdoor_param} and let $P(G_{ij}|D)$ denote the posterior probability of the adjustment set $G_{ij}$ with respect to a predefined class of adjustment sets.
The mean of the mixture distribution in Equation \eqref{eq::posterior_pi} can then be computed elementwise as:
\begin{align*}
&E[\pi_{x_j|x_i}|\, D] = \\
&\qquad \frac{\alpha_{x_j}  + N_{x_j}}{\alpha_0 + N}\times  \sum_{G_{ij}:j\in G_{ij}} P(G_{ij}|\, D) \; + \\ & \sum_{G_{ij}:j\notin G_{ij}}P(G_{ij}|\, D) \sum_{x_{G_{ij}}}\frac{\alpha_{x_j\vert  x_i, x_{G_{ij}}} + N_{x_j\vert  x_i, x_{G_{ij}}}}{\sum_{x_j}\alpha_{x_j\vert  x_i, x_{G_{ij}}} + N_{x_j\vert  x_i, x_{G_{ij}}}}\frac{\alpha_{ x_{G_{ij}}} + N_{ x_{G_{ij}}}}{\alpha_0 + N},
%\end{align*}
%\begin{align*}
%&E[\pi_{x_j|x_i}|\, D] = P(j\in G_{ij}|\, D) \times\tilde\alpha_{x_j}^D+ \sum_{G_{ij}:j\notin G_{ij}}P(G_{ij}|\, D) \sum_{x_{G_{ij}}}\tilde\alpha^D_{x_j\vert x_i, x_{G_{ij}}}\tilde\alpha^D_{x_{G_{ij}}}
\end{align*}
where $\alpha_0$ is the equivalent sample size and $N$ the number of samples in $D$.
%$$P(j\in G_{ij}|\, D) = \sum_{G_{ij}:j\in G_{ij}} P(G_{ij}|\, D).$$
\end{corollary}

\subsubsection{Posterior over backdoor adjustment sets \label{sec::posterior-parents}}
In estimating the posterior over IPTs in \eqref{eq::posterior_pi}, computing the posterior $p(G_{ij}|\, D)$ over the adjustment sets is computationally the most challenging task. For the original BIDA method, \citet{Pensar2020} proposed an exact algorithm for computing of the posterior probabilities of parent sets in smaller networks, with up to about 25 nodes. However, in larger domains one needs to resort to approximate methods. To this end, Markov Chain Monte Carlo (MCMC) is the most common approach for Bayesian structure learning. By generating a sample of graphs from the posterior distribution, expectations of given network features can be estimated by averaging over the sample. Here, our target is the posterior distribution over adjustment sets, $G_{ij}$, of some predefined class, which can be estimated as 
$$
P(G_{ij}|D) \approx \frac{1}{M}\sum_{m=1}^M \mathbf{1}\{G_{ij} = G_{ij}^{(m)}\}, 
$$
where $M$ is the sample size and $\mathbf{1}\{G_{ij} = G_{ij}^{(m)}\}$ equals one if the adjustment set in the $m$:th sampled DAG, $G_{ij}^{(m)}$, equals $G_{ij}$ and zero otherwise.

The MCMC approach does not only enable the parent-based BIDA procedure to be scaled up, it also enables the use of alternative adjustment sets, since the existing state-of-the-art algorithms generates complete DAGs. In particular, by conditioning on the existence of causal paths in the sampled graphs, causal effects can be set to exactly zero, while still avoiding the need to perform inference in the full model. 

Multiple MCMC-based variants have been proposed for the purpose of learning (causal) DAGs \citep{Friedman2003,Kuipers2017,Viinikka2020,kuipersEfficientSamplingStructure2022}. 
In this work, we focus on the recent algorithm by \citet{kuipersEfficientSamplingStructure2022}, which combines partition-MCMC with constraint-based structure learning.
The partition-MCMC scheme operates on the space of node partitions. As DAGs can be uniquely assigned to node partitions, partition-MCMC does not suffer from the bias of order-MCMC which operates over node orderings \citep{Friedman2003}. For improved scalability, the hybrid algorithm limits the initial search space of partition-MCMC to a skeleton that can be efficiently obtained, for example, with a constraint-based method such as the PC-algorithm \citep{Spirtes1993}. The search space is then allowed to expand iteratively, to correct for errors where true edges have initially been omitted from the skeleton.

\subsection{Posterior of the causal effect \label{sec::posterior_causal_effect}}

The posterior over intervention distributions, $p(\pi_{ij}|D)$, in \eqref{eq::posterior_pi} can now be used to predict the (causal) effect of interventions. As proposed in Section \ref{sec::JSD}, we define the causal effect $\tau_{ij}$ through the Jensen-Shannon-divergence (JSD) \eqref{eq::JSD} as a function of the IPT:
$$
\tau_{ij} = \frac{1}{r_i}\sum_{x_i}\sum_{x_j}\pi_{x_j|x_i} \log\frac{\pi_{x_j|x_i}}{\frac{1}{r_i}\sum_{x_i}\pi_{x_j|x_i}}. 
$$
While there is in general no closed form expression for the implicit posterior distribution $p(\tau_{ij}|D)$, it is straightforward to sample from the exact distribution to obtain Monte Carlo approximations of a target posterior quantity, for example, the posterior mean. 
In Appendix B in the supplementary material\citesupp, we show some example causal effect posteriors under a small toy scenario, illustrating the large sample behavior and the potential multi-modality discussed in Section \ref{section::posterior-of-intervention-probabilities}.

\subsubsection{Ranking of causal effects}
The main motivation behind the BIDA procedure is to identify strong causal relationships in a system and, as such, a desired output of the procedure is informative rankings of the causal effects. One advantage of the Bayesian approach is that it provides a posterior distribution over each effect, based on which various rankings of cause-effect pairs can be based. 
In this work, we consider two different ranking strategies in which we rank the causal effects according to their posterior mean values and the posterior mean ranks. 

The posterior mean rank of the causal effects is defined as follows. Based on the matrix of causal effects, $\tau$, we define the $n\times n$ rank matrix $R(\tau) = \{R_{ij}(\tau)\}$:
\begin{align}
    R_{ij}(\tau) = \sum_{l=1}^n\sum_{k=1}^n 1(\tau_{ij}\geq \tau_{lk}).\footnotemark  \label{eq::rank_tau}
\end{align}
The posterior mean rank of the causal-effect pair $(i, j)$ is then given by:
$$E[R_{ij}(\tau)|D] = \sum_{l=1}^n\sum_{k=1}^n P(\tau_{ij}\geq\tau_{lk}|\,D),$$
and it can in practice be approximated from a sample of causal effect matrices, $t^{(1)}, ..., t^{(M)}$, generated from $p(\tau|D)$, as the average rank:
$$ E[R_{ij}(\tau)|D] \approx \frac{1}{M}\sum_{m = 1}^M R_{ij}(t^{(m)}).$$
The posterior mean rank is proportional to the probability that $\tau_{ij}$ is larger than or equal to the causal effect $\tau_{lk}$ of a randomly chosen pair $(l, k)$. 
Compared to the mean value of the causal effect, the mean is more reflective of where the posterior distributions have the most mass. 
The mean rank will therefore typically rank pairs with right-skewed posteriors lower, where the confidence in large effects is relatively low. 
Such distributions arise, for example, when the estimated effect is positive only in a few low-probability structures.

\footnotetext{Note that, by this definition, the same rank is assigned to elements in $\tau$ with identical values. The rank of every zero effect, for example, equals the total number of zero effects in $\tau$. Except from those equal to zero, we do not expect any pair of sampled effects to be exactly equal.}

\onlyifstandalone{\bibliographystyle{apalike}\bibliography{\rootdir{library}}}

%% file: sections/numerical_experiments.tex
\section{Numerical experiments \label{sec::num_exp}}
We evaluate our proposed BIDA method in a simulation study, where the aim is to estimate the causal effects from observational data sampled from ground-truth Bayesian networks.
We focus on two central objectives when estimating causal effects under an unknown structure: the accuracy of the estimator and its ability to discover strong effects, as measured by the mean squared error (MSE) and the average under the precision-recall curve (AUC-PR), respectively.
For comparison purposes, we included two versions of IDA, namely the original version \citep{Maathuis2009} and the optimal IDA \citep{Witte2020}, both adapted to the categorical setting, as well as the Bayesian model averaging approaches presented in \cite{moffaUsingDirectedAcyclic2017} (Bestie) and \cite{castellettiJointStructureLearning2024} (CCDV). 
However, the two Bayesian approaches are in their current implementations only applicable to binary data. 
Therefore, we performed two simulation experiments, one for the binary setting and one for the general categorical setting. In the first experiment, we used randomly generated networks. In the second experiment, we used 9 benchmark networks from the bnlearn repository \citep{scutariLearningBayesianNetworks2010}. Details on the implementation of the considered methods and our chosen evaluation criteria is included in the supplement material to this paper\citesupp, which also includes additional simulation results.

\subsection{Binary setting}

\subsubsection{Simulation setup}
We randomly generated 30 DAGs over $n = 10, 20$ nodes with an expected number of neighbours equal to 4 and, for each DAG, the associated model parameters such that each $\theta_{X_i|x_{G_i}}$ was drawn from a uniform Dirichlet distribution $(\alpha_{x_i\vert x_{G_i}} = 1)$.
To align with the R implementations of Bestie and CCDV, we target in this binary setting the average treatment effect (ATE) for every cause-effect pair $(i, j)$, as defined in equation \eqref{eq::ATE},  rather than the JSD.
We used forward sampling to obtain Monte Carlo approximations of the true ATEs. For each network, we sampled data sets of three different sample sizes: $N = 300, 1 000, 3 000$, which we used to estimate the ATEs. As point estimates, we used the posterior means for the Bayesian approaches and the average of the returned set of point estimates for IDA. We applied BIDA with parent and minimal parent sets, as we found minimal parent sets to be our preferred choice of adjustment set (we discuss this in more detail in the next section). Furthermore, for Bestie and CCDV, we sampled DAGs and computed for every DAG the ATEs from the (conditional) posterior mean values of the model parameters, rather than a posterior sample of parameters.\footnote{We observed that, in our simulation setting, the accuracy of the point estimates of CCDV were considerably lower when both DAGs and parameters were sampled, as in their original version, see Figure D.1 in the supplement.}
Since the {CCDV} implementation took several hours to run with $n=10$, we did not consider this method for $n=20$. Further details on the implementation of each method are included in Appendix C in the supplement.
\begin{figure}[t!]
    \centering
    \includegraphics[width = \linewidth]{\rootdir{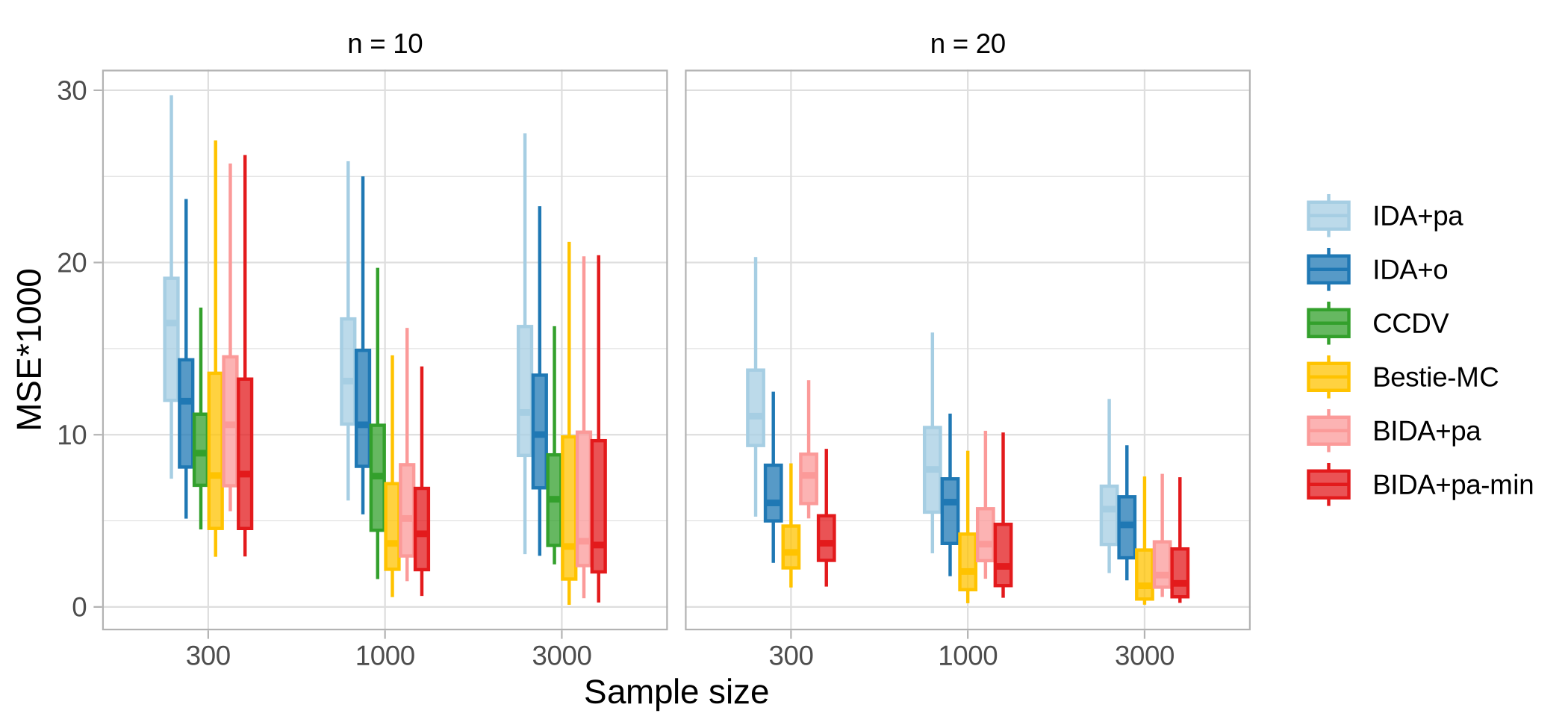}}
    \caption{
    The accuracy of BIDA with parent sets (BIDA+pa) and minimal parent sets (BIDA+pa-min), as measured by the mean squared error (MSE) between the true and estimated ATE. Additional methods included are the original IDA (IDA+pa), the optimal IDA (IDA+o) and two Bayesian model averaging approaches (Bestie and CCDV). The boxplot shows the distribution over 30 simulation runs, where data sets of different sizes were sampled from a random network of size $n = 10$ (left figure) and $n=20$ (right figure). Outliers are not shown. The CCDV is not included in the $n=20$ setting due to long run times. \label{fig::sim_binary_mse}}
\end{figure}

\subsubsection{Results}
In terms of estimating the causal effects, here defined by the ATE, the accuracy of BIDA is comparable to that of the best performing approach, Bestie, when minimal parent sets are used for adjustment in the backdoor formula (Figure \ref{fig::sim_binary_mse}). With parent sets, BIDA is not as accurate as Bestie, especially not for the smaller sample sizes.
As BIDA and Bestie are given the same sample of DAGs, they differ only in how they build a posterior over the causal parameters of interest: BIDA returns a local approximation of the target posteriors based on the backdoor formula, while Bestie makes inference in the full model. When we use minimal parent sets, BIDA sets the ATE equal to zero for all cause-effect pairs that are not causally connected in a given DAG. As a result, the point mass at zero is the same in the BIDA and Bestie posteriors.

BIDA and Bestie are also the top performers in terms of discovering strong causal relationships, as measured by the area under the precision recall-curve (Figure D.2 in the supplement). Here ``strong'' causal relationships are defined as those for which the magnitude of the true ATE lie in the top 20 percent across the non-zero effects in the network. 

To compare the structure learning procedures involved in each method, we evaluated their precision in predicting causal relationships (that is, positive effects). Specifically, we compared the ancestor relation probabilities (ARPs)\footnote{The ARPs are estimated directly from the MCMC sample of graphs. For each cause-effect pair $(i,j)$, we simply compute the fraction of sampled DAGs where $i$ is an ancestor of $j$.} of the DAGs sampled with partition-MCMC and CCDV with those implied by the CPDAG infered by the PC algorithm. The Bayesian approaches achieve the highest AUC-PR scores, suggesting that they more accurately recovers the true causal structure. The differences between partition-MCMC and CCDV are minor (Figure D.3 in the supplement).

As expected, our proposed BIDA method is computationally more efficient compared to the two existing Bayesian approaches. For the largest network ($n = 20$), conditional on a sample of DAGs, it takes BIDA a couple of seconds to compute point estimates, while Bestie requires a couple of minutes (Table D.1 in the supplement). We also note a decrease in run time for these methods when the sample size increases. The reason is that the MCMC chain tends to concentrate around some high scoring node partitions as the sample size increases, leading to a decrease in the number of unique DAGs in which inference has to be conducted.

\subsection{General categorical setting}
\subsubsection{Simulation setup\label{sec::sim-setup}}
For the general categorical setting, we studied the performance of our proposed BIDA method and non-Bayesian IDA on 9 discrete Bayesian networks from the bnlearn repository \citep{scutariLearningBayesianNetworks2010}. The number of nodes in the considered networks varies from 8 to 76. Other key characteristics are shown in Table E.1 in the supplementary material to this paper.
For all networks, we used forward sampling to obtain Monte Carlo approximations of the true causal parameters of interest, namely the marginal intervention probabilities, $\pi$, and the pairwise causal effect $\tau$, which are now defined by the JSD \eqref{eq::JSD}.
Next, we sampled 30 data sets of three different sample sizes, $N = 300, 1 000, 3 000$, from each network and estimated these parameters. 
We focus on three networks when we present the results of the simulation study in the following sections: Child ($n=20)$, Alarm ($n = 37)$ and Win95pts $(n = 76)$. However, results for all 9 networks are included in Appendix E in the supplement.

%\subsubsection{Results}
\subsubsection{Results: Choice of adjustment set}
% FIG: COMPARING ADJUSTMENT SETS
\begin{figure}[t!]
    \centering
    \includegraphics[width = \linewidth]{\rootdir{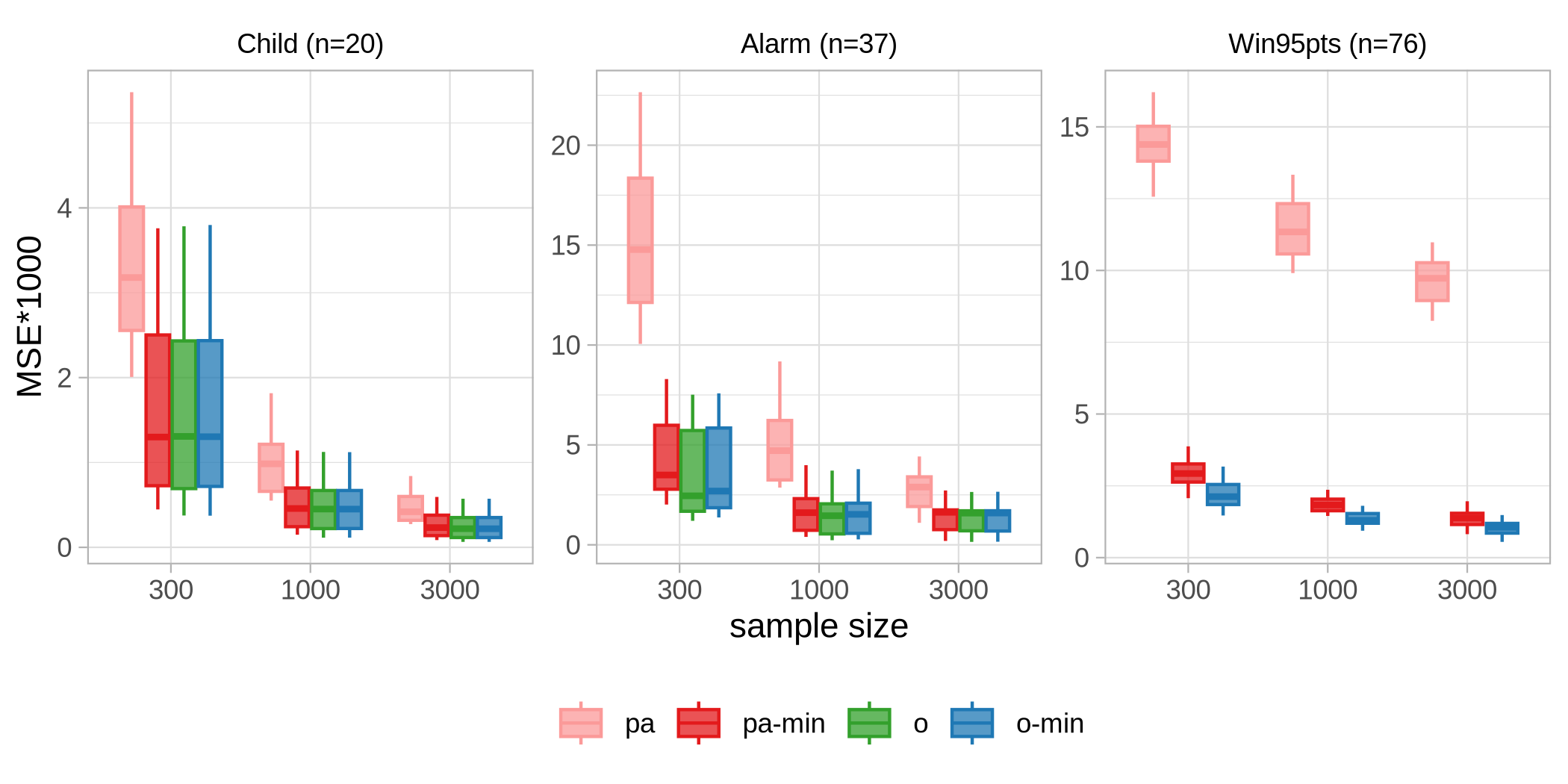}}
    \caption{The accuracy of BIDA paired with four different adjustment sets, as measured by the mean squared errors (MSE) between the true and estimated causal effects, $\tau$. The adjustment sets considered are parent sets (pa), minimal parent sets (pa-min), o-sets (o) and minimal o-sets (o-min). Each subfigure corresponds to one network and the boxplot shows, for each sample size, the distribution over 30 data sets sampled from the network. Outliers are not shown. 
\label{fig::mse_adjust}}
\end{figure}
We first study how the accuracy of the BIDA procedure is affected by the choice of adjustment set. 
Although our primary interest is in the relative accuracy of these sets when the true underlying DAG is unknown, we also compare their accuracy under a known DAG. 

Our results show that the accuracy of BIDA can be greatly increased by replacing parent sets with either o-sets or minimal sets, both when the true causal DAG is unknown (Figure \ref{fig::mse}) and when it is assumed known (Figure E.1 in the supplement). 
The differences in accuracy between the minimal parent set, the minimal and the full o-set are smaller, reflecting the fact that they agree on all zero effects. 
Still, the two o-sets are typically the most accurate. Across all 9 networks there was only one setting where one of these adjustment sets did not achieve the lowest MSE (Figure E.1 in the supplement). 
Comparing the o-set and the minimal o-set for all the small and medium sized networks ($n < 50$), the difference is negligible under an unknown DAG. Under the true DAG neither set dominates the other. To reduce computational costs, we chose therefore to not apply the BIDA method with the full o-set on the three largest networks. 

Overall, these results suggest a significant improvement in accuracy from including a search for causal paths and setting effects to exactly zero in structures where no such path exists. Also, they give support to the theoretical arguments for confounding adjustment with covariates close to the effect rather than the cause \citep{witteCovariateSelectionStrategies2019,Witte2020, henckelGraphicalCriteriaEfficient2022}. The benefit in terms of estimation accuracy is, naturally, more pronounced under a known DAG.
The improved accuracy of the (minimal) o-set does come with an increased computational cost, at least in the larger networks. 
Computing o-sets and minimal sets in a sample of DAGs requires substantially higher computational effort compared to the parent sets. 
Additionally, in larger networks the o-sets (and minimal o-sets) can be very large, and computing the posterior hyperparameters  and sampling from the posterior then more demanding. 
On the other hand, sampling from the posterior is also less demanding as more estimates are set to zero.
We observed that the latter effect dominated in small networks, for which the parent-based BIDA had the highest run times (see Appendix E.3 in the supplement).
The average size of the adjustment sets is compared in Appendix E.2.

Balancing statistical and computational efficiency, the minimal parent set is our preferred adjustment set, and we will focus on results from pairing this set with our BIDA procedure in the following.

% FIGURE: MSE-BIDA-VS-IDA
\begin{figure}[t!]
    \centering
    \includegraphics[width = \linewidth]{\rootdir{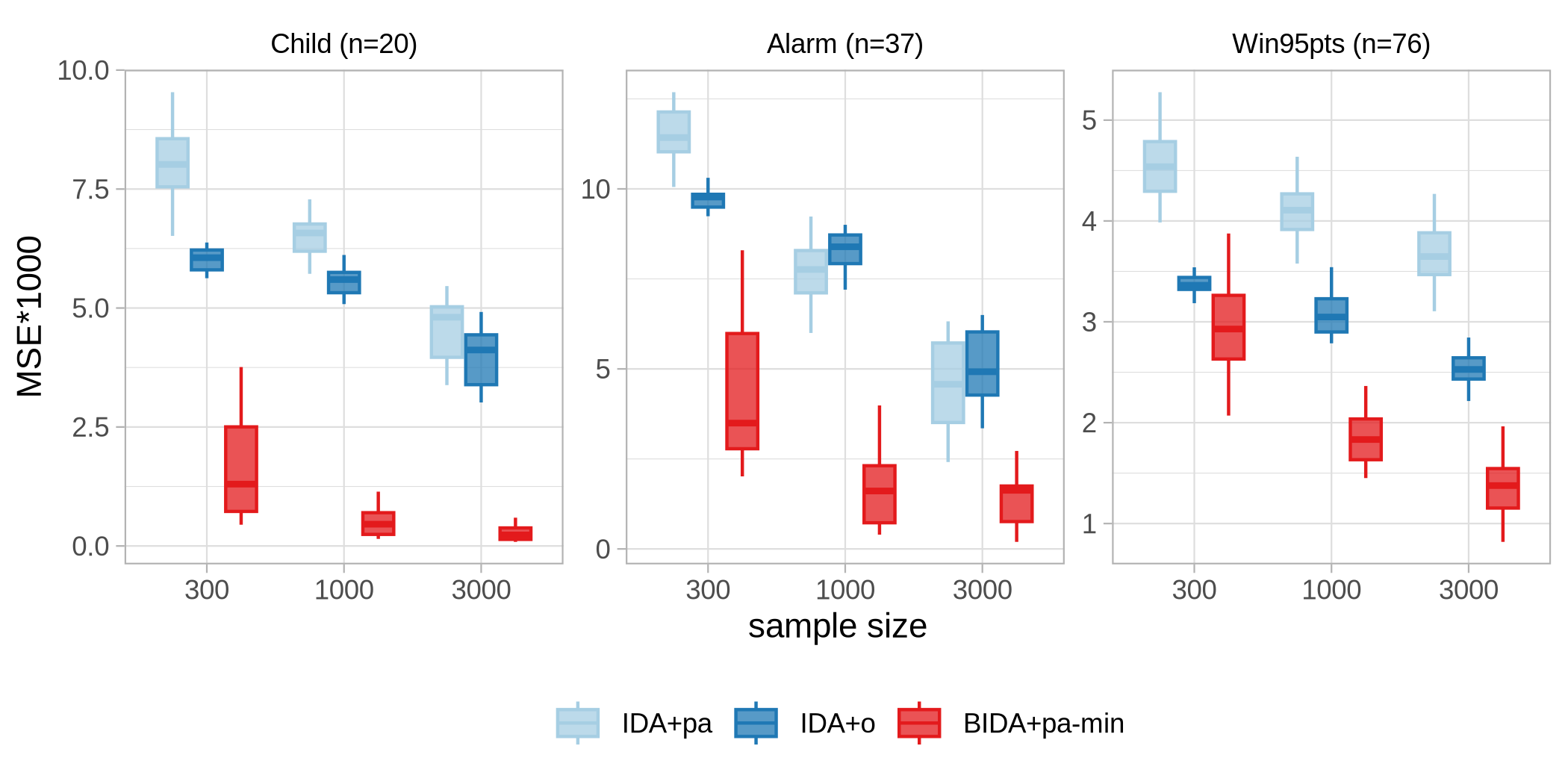}}
    \caption{The accuracy of BIDA (BIDA+pa-min) compared to the original IDA (IDA+pa) and the optimal IDA (IDA+o), as measured by the mean squared errors (MSE) between the true and estimated causal effects $\tau$.  Each subfigure corresponds to one network and the boxplot shows, for each sample size, the distribution over 30 data sets sampled from the network. Outliers are not shown. \label{fig::mse}}
\end{figure}

\subsubsection{Results: Comparison with the IDA algorithm}

%% AUC-PR
\begin{figure}[t!]
    \centering
    \includegraphics[width = \linewidth]{\rootdir{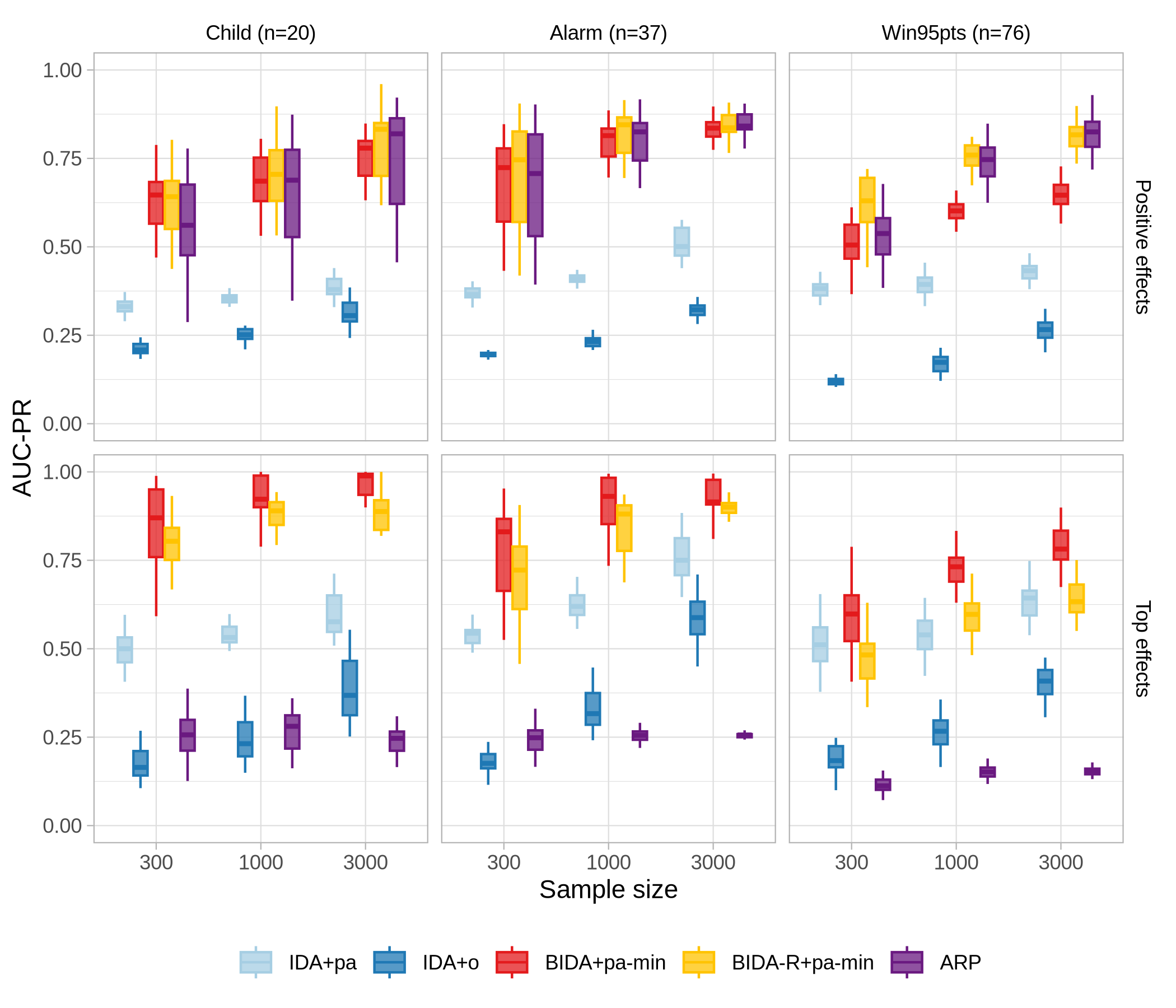}}
    \caption{The accuracy in predicting strong effects, as measured by the area under the precision-recall curve (AUC-PR), using the posterior mean values (BIDA+pa-min), posterior mean ranks (BIDA-R+pa-min), original IDA (IDA+pa), optimal IDA (IDA+o) and ancestor relation probabilities (ARP). 
    Each subfigure corresponds to one network and the boxplot shows, for each sample size, the distribution over 30 data sets sampled from the network. Outliers are not shown. 
    In the top row the true positives are all non-zero effects, in the bottom row only those in the top 20th percentile.\label{fig::aucpr}}
\end{figure}

Our results show that BIDA overall produces more accurate point estimates compared to those of parent-based IDA and optimal IDA.
For the three networks considered in Figure \ref{fig::mse}, BIDA clearly attains the lowest MSE for all sample sizes. 
When considering all networks, however, there are a few cases where our method struggles. In particular, its relative performance is rather poor on the \emph{Sachs} and \emph{Water2} network (see Figure E.2 in the supplement).
We trace these poor results back to the structure learning part of the procedure, observing that partition-MCMC has trouble mixing given data sampled from these two networks. The chain visits very few structures and is likely not a good approximation of the posterior distribution. Consequently, the resulting point estimates are not very accurate. This highlights the importance of obtaining a representative sample from the posterior over the adjustment sets for the BIDA method.

Among the two IDA variants, the optimal IDA is typically the most accurate.
It reaches a very low MSE in some of the networks, but then together with the reference based on marginal probabilities (Figure E.2 in the supplement). 
This indicates that the PC algorithm returns very sparse structures and that most of the true effects are either zero or very small in these networks.

In addition to the point estimates of the causal effect matrix $\tau$, we compared estimates of the intervention probabilities, $\pi$. 
Accurate estimates of these parameters are relevant if one wants to consider alternative contrasts of the intervention distributions, such as, for example, the average treatment effect (ATE). The relative accuracy of the methods in estimating $\pi$ is very similar to that observed for the causal effects $\tau$, as shown in Figure E.3 in the supplement.

In terms of discovering strong causal relationships, we find that BIDA most often performs better than the non-Bayesian IDA variants (Figure \ref{fig::aucpr}).
In predicting the mere existence of causal relationships (that is, positive effects), we obtain a higher AUC-PR using the posterior mean rank of the causal effects (see Section \ref{sec::posterior_causal_effect}), compared to the posterior mean values. The performance of the former is close to that of the ARPs estimated directly from the MCMC sample of graphs, which is a useful baseline in this case. 
However, when we increase the threshold that defines the true strong effects, as in the bottom panels of Figure \ref{fig::aucpr}, the highest AUC-PR is achieved by the posterior mean values.
The ARP is, as expected, clearly not good at separating large positive effects from smaller, as it does not explicitly take the magnitude of the effect into account.

We see the same pattern across most of the networks (Figure E.4 and Figure E.5 in the supplement). 
The relative good performance of the posterior mean ranks under both thresholds suggests that it is more robust to changes in the threshold, compared to the posterior mean values. 
Still, neither outperforms the IDA algorithm in every setting we consider. Again, their relative performance is poor on the \emph{Sachs} and \emph{Water2} network, as a result of the mixing issue. However, we see that the absolute performance across methods is rather poor for these particular networks. 

Comparing the two IDA-algorithms, the original parent-based method is the most accurate in predicting positive and large effects, despite being the least accurate of the two in terms of MSE. 
As discussed in \citet{Maathuis2009}, the local parent-based version of the IDA algorithm is more robust to errors in the inferred structure, in the sense that the lack of causal paths between a pair of nodes does not imply that the estimated effect is zero.
Interestingly, we do not see the same disadvantage of conditioning on causal paths with our BIDA procedure. Compared to the fully local parent based version, we most often obtain both more accurate point estimates and more accurate predictions of strong effects when we include a search for causal paths and set effects to zero in structures where no such paths exist. Overall, by considering more than a single equivalence class, the Bayesian procedure is better equipped to account for inherent uncertainty in the structure learning step.

\onlyifstandalone{\bibliographystyle{apalike}\bibliography{\rootdir{library}}}

%% file: sections/discussion.tex
\section{Discussion and conclusion \label{sec::discussion}} 
We have presented a Bayesian approach (BIDA) for estimation of causal effects from observational data under an unknown causal structure. Most of the existing procedures developed for this setting assume a linear Gaussian \citep{Maathuis2009, Witte2020, Pensar2020, Viinikka2020, Castelletti2021} or a binary model \citep{moffaUsingDirectedAcyclic2017, castellettiJointStructureLearning2024}. Here, we have assumed the system is composed of a set of categorical variables, extending the applicability of the original BIDA method. In addition, to scale up the procedure, we employ MCMC to generate a sample of complete DAGs from the graph posterior, which also enables us to study various classes of adjustment sets.

We studied the performance of our proposed BIDA method in a simulation experiment, estimating all pairwise causal effects both in synthetic networks and in 9 discrete Bayesian networks from the bnlearn repository.
Comparing the accuracy of BIDA across four different classes of backdoor adjustment sets (the full parent set, the o-set and the minimal versions of both), we found that the o-sets gave the most accurate estimates, in line with asymptotic theory and previous simulation studies \citep{witteCovariateSelectionStrategies2019, Witte2020}. For the considered context, we preferred the minimal parent sets for its lower computational costs, as the full o-set can grow very large when the number of nodes is increased. Compared to two full Bayesian model averaging (BMA) approaches implemented for the binary setting, we demonstrated that the BIDA behaved similarly to the best of these approaches, at a lower computational cost. In the general categorical setting, we demonstrated that BIDA overall produces more accurate point estimates and rankings over the causal effects, compared to two non-Bayesian IDA alternatives.  Nonetheless, the performance of our method was rather poor in some of the networks, in which partition-MCMC suffered from mixing issues. Further developments and improvements of methods for sampling DAGs would be beneficial for the BIDA approach, both from a computational and statistical perspective.

In terms of the considered IDA variants, the optimal IDA was the best in terms of accuracy, while the original procedure was better at predicting strong effects. The main reason for this is that the PC algorithm often returned very sparse structures, resulting in many zero estimates. The majority of the true pairwise causal effects were indeed zero in all networks, biasing the overall accuracy towards more conservative estimates. 
%Obviously, a good method for assessing the underlying causal structure is crucial also for the BIDA procedure. 

In this work, we defined the causal effect as the Jensen-Shannon divergence (JSD) between intervention distributions, which allowed us to summarize, as a single quantity $\tau_{ij}$,  the effect of all possible interventions on $X_i$ on every outcome of $X_j$ for the general categorical setting. Equal to the mutual information between a random intervention and the outcome, it has an intuitive information-theoretic interpretation, although, from a statistical point of view, it can be challenging to estimate.  
However, the proposed procedure is not tied to a specific causal effect, but can also be used with other contrasts of intervention distributions. 

To scale up the procedure even further, one possibility is to simplify the representation of the posterior distribution over intervention probabilities through some form of approximation. Sampling these parameters from their exact posterior distributions is computationally demanding, especially if the considered variables are of high cardinality or the adjustment sets are large. While there is no closed-form expression of the exact posterior distribution in the general case, we have derived certain moments of the distribution on which such approximations could be based.
Another interesting extension would be to combine the procedure with more sophisticated ranking models for the purpose of discovering the top ranked cause-effect pairs. 
In a third and last direction, it would be natural to delve deeper into and improve the critical structure learning part of the procedure. In addition to developing new methods for sampling DAGs, an interesting future direction is to consider model structures that are more flexible than standard discrete Bayesian networks \citep{friedmanLearningBayesianNetworks1998, Pensar_2014,Hyttinen2018}.

As a final note, we remind that the causal interpretation of the proposed method relies on strong and untestable assumptions. Even if these hold, without knowledge of the causal model, causal effects can not be uniquely inferred from non-experimental data alone. The main motivation of the BIDA procedure is to guide and inform future experiments by picking out plausible strong causal relationships, which then need to be verified experimentally. In this respect, it is precisely the ability to consistently rank the true cause-effect pairs that is important. For that purpose, our simulation experiments demonstrate that the BIDA procedure is a relevant tool in the causal inference toolbox, and, by our extensions, applicable to a wider set of problems.

%% file: sections/appendix-moments.tex
\section{Moments of the distribution over intervention probabilities \label{appendix::mom}}

As in Theorem 1, assume that the random vectors $\theta_{Y|x^1, z^1}, \ldots, \theta_{Y|x^{r_x}, z^{r_x}}$ and $\theta_Z$ are mutually independent and that each vector is Dirichlet distributed:
$$
\begin{aligned}
    p(\theta_{Y|x, z}) &= \text{Dirichlet}(\alpha_{y^1\vert x,z}, \ldots, \alpha_{y^{r_y}\vert x,z}), \text{ for each } (x,z), \\
    p(\theta_{Z}) &=\text{Dirichlet}(\alpha_{z^{1}}, \ldots, \alpha_{z^{r_z}}), 
\end{aligned}
$$

\noindent such that $\alpha_{z} = \sum_x\sum_y \alpha_{y\vert x,z}$ for all $z$, and consider the random variables that result from the linear combinations: 
$$\pi_{Y|x} = \sum_z\theta_{y|\, x,z}\theta_{z}, \text{ for each } x. $$
The exact distribution of $\pi_{Y|x}$ can, in the general case, not be expressed in closed-form.  In the following, we derive expressions for the moments given in Theorem 1: (i) the mean $E[\pi_{y|x}]$, (ii) the covariance $Cov(\pi_{y|x}, \pi_{y'|x})$, and (iii) the covariance between parameter pairs of two different linear combinations, $Cov(\pi_{y|x}, \pi_{y'|x'})$, where $x \neq x'$.

First, we note that the linear combination of Dirichlet vectors are shown to be Dirichlet distributed under the constraint $\alpha_{z} = \alpha_{x, z}$ 
\citep{Homei2021}. Also, the marginal distribution of the scaled vectors $\theta_{Y|x, z^1}\theta_{z^1}, \ldots, \theta_{Y|x, z^{r_z}}\theta_{z^{r_z}}$ has, for each $x$, density
\begin{align*}
p(\theta_{Y|x, z^1}\theta_{z^1}, & \ldots, \theta_{Y|x, z^{r_z}}\theta_{z^{r_z}}) = \\
&\Gamma(\alpha)\prod_{z} \frac{\Gamma(\alpha_{x, z})}{\Gamma(\alpha_z)}\left( \sum_y\theta_{y|x,z}\theta_z \right)^{{\alpha_z}-\alpha_{x, z}}\prod_{y}\left[\frac{(\theta_{y|x,z}\theta_z)^{\alpha_{y, x,z}-1}}{\Gamma(\alpha_{y, x,z})}\right],    
\end{align*}
where $\alpha = \sum_z\alpha_z$ and $\alpha_z = \sum_x \alpha_{x, z}$ and $\alpha_{x,z} = \sum_y \alpha_{y\vert x, z}$. 
The above density characterizes the so-called grouped Dirichlet distribution \citep{ngDirichletRelatedDistributions2011}. 
Again, for the special case where $\alpha_{z} = \alpha_{x, z}$,  the above density is a Dirichlet and, by the aggregation properties of this distribution, the linear combination $\pi_{Y|x}$ is then Dirichlet distributed. 
For the general case, the selected moments of the joint distribution are derived as follows.
\vspace{1em}
\noindent (i) The mean of each element in $\pi_{Y|x}$ is given simply by
\begin{align*}
E[\pi_{y|x}] &= E[\sum_z \theta_{y|\, x,z}\theta_{z}] = \sum_z E[ \theta_{y|\, x,z}]E[\theta_{z}] =\sum_z  \tilde\alpha_{y|x, z}\tilde\alpha_z ,
\end{align*}
where the Dirichlet means are denoted as: $\tilde\alpha_{y\vert x,z} = \alpha_{y \vert x,z}/\alpha_{x,z}$ and $\tilde\alpha_{z} = \alpha_{z}/\alpha$ with $\alpha = \sum_z\alpha_z$ and $\alpha_{x,z} = \sum_y\alpha_{y\vert x,z}.$

\vspace{1em}
\noindent (ii) The covariance between a pair of parameters $(\pi_{y|x}, \pi_{y'|x})$ is
\begin{align*}
Cov(\pi_{y|x}, \pi_{y'|x}) 
&= E[\pi_{y|x}\pi_{y'|x}] - E[\pi_{y|x}]E[\pi_{y'|x}].
\end{align*}
 
\noindent The independence assumptions implies that the first term can be written as:
\begin{align}\label{eq::firstprodmom}
E[\pi_{y|x}\pi_{y'|x}] 
&=\sum_z\sum_{z'} E[\theta_{y|x, z}\theta_z\theta_{y'|x, z'}\theta_{z'}] \notag \\
&=\sum_z \left( E[\theta_{y|x, z}\theta_{y'|x, z}] E[\theta_z\theta_{z}] + \sum_{z'\neq z} E[\theta_{y|x, z}]E[\theta_{y'|x, z'}] E[\theta_z\theta_{z'}]\right).
\end{align}

\noindent As $\theta_{y|x, z}$ and $ \theta_{y'|x, z}$ are elements of the same Dirichlet vector, their product has expected value:
\begin{align*}
E[\theta_{y|x, z}\theta_{y'|x, z}] 
&= Cov(\theta_{y|x, z}\theta_{y'|x, z}) + E[\theta_{y|x, z}]E[\theta_{y'|x, z}] \\
&=
\frac{\delta_{yy'}\tilde\alpha_{y|x, z}-\tilde\alpha_{y|x, z}\tilde\alpha_{y'|x, z}}{\alpha_{x, z}+1} + \tilde\alpha_{y|x, z}\tilde\alpha_{y'|x, z}\\
&=\frac{\tilde\alpha_{y|x, z}(\delta_{yy'}+\alpha_{x,z}\tilde\alpha_{y'|x, z})}{\alpha_{x,z}+1}.
\end{align*}
where $\delta_{yy'}$ equals 1 if $y = y'$ and zero otherwise. Similarly, we have
\begin{align*}
   E[\theta_z\theta_{z'}] = \frac{\tilde\alpha_z(\delta_{zz'}+\alpha\tilde\alpha_{z'})}{\alpha+1} = \frac{\tilde\alpha_z(\delta_{zz'}+\alpha_{z'})}{\alpha+1},
\end{align*}
where in the last equation we have used that $\alpha\tilde\alpha_{z'} = \alpha_{z'}$.
Using the above Dirichlet moments, we can write the summands in Equation \eqref{eq::firstprodmom}: 
\begin{align*}
 E[\theta_{y|x, z}\theta_{y'|x, z}] E[\theta_z\theta_{z}] 
 &= \frac{\tilde\alpha_{y|x, z}(\delta_{yy'}+\alpha_{x, z}\tilde\alpha_{y'|x, z})}{\alpha_{x, z}+1}\cdot \frac{\tilde\alpha_z(1+\alpha_z)}{\alpha+1}\\
 &=\frac{\tilde\alpha_{y|x, z}\tilde\alpha_z}{(\alpha+1)(\alpha_{x, z}+1)}\left(\delta_{yy'}(1+\alpha_z)+\alpha_{x, z}\tilde\alpha_{y'|x, z}(1+\alpha_z)\right), 
\end{align*}
and
\begin{align*}
\sum_{z'\neq z} E[\theta_{y|x, z}]E[\theta_{y'|x, z'}] E[\theta_z\theta_{z'}] 
&= \frac{\alpha}{\alpha+1}\tilde\alpha_{y|x,z}\tilde\alpha_z\sum_{z'\neq z}\tilde\alpha_{y'|x,z'}\tilde\alpha_{z'}  \\
&= 
\frac{\alpha}{\alpha+1}\tilde\alpha_{y|x,z}\tilde\alpha_z\left(E[\pi_{y'|x}]-\tilde\alpha_{y'|x,z}\tilde\alpha_{z} \right). 
\end{align*}
If we collect the terms that involve $\tilde\alpha_{y|x, z}\tilde\alpha_{y'|x, z}\tilde\alpha_z/(\alpha+1)$, that is
$$
\frac{\tilde\alpha_{y|x, z}\tilde\alpha_{y'|x, z}\tilde\alpha_z}{\alpha+1}
\left(\frac{\alpha_{x, z}(1+\alpha_z)}{\alpha_{x, z}+1} - \alpha\tilde\alpha_z \right) = \frac{\tilde\alpha_{y|x, z}\tilde\alpha_{y'|x, z}\tilde\alpha_z}{\alpha+1}\left(\frac{\alpha_{x, z}-\alpha_z }{\alpha_{x, z}+1}\right),
$$
and use $\sum_z\tilde\alpha_{y|x,z}\tilde\alpha_zE[\pi_{y'|x}] = E[\pi_{y|x}]E[\pi_{y'|x}]$, the expectation in \eqref{eq::firstprodmom} can be written as
\begin{align*}
  &E[\pi_{y|x}\pi_{y'|x}] = \\
  &\frac{1}{\alpha+1}\left(\sum_z \frac{\tilde\alpha_{y|x, z}\tilde\alpha_z}{\alpha_{x, z}+1}\left(\delta_{yy'}(1+\alpha_z)+\tilde\alpha_{y'|x, z}(\alpha_{x, z}-\alpha_z)\right)  + \alpha E[\pi_{y|x}]E[\pi_{y'|x}]  \right). 
\end{align*}
Finally, the covariance between $\pi_{y|x}$ and $\pi_{y'|x}$ can then be expressed:
$$ 
\begin{aligned}
&Cov(\pi_{y|x}, \pi_{y'|x'}) = \\
&\quad  \frac{1}{\alpha+1}\left(\sum_z\frac{\tilde\alpha_{y|x, z}\tilde\alpha_z}{\alpha_{x, z}+1}\left(\delta_{yy'}(1+\alpha_z)+(\alpha_{x, z}-\alpha_{z})\tilde\alpha_{y'|x, 
z}\right) - E[\pi_{y|x}]E[\pi_{y'|x}]\right).
\end{aligned}
$$

\vspace{1em}
\noindent (iii) The covariance between a pair of parameters $(\pi_{y|x}, \pi_{y'|x'})$ with $x\neq x'$ is
\begin{align*}
Cov(\pi_{y|x}, \pi_{y'|x'}) 
&= E[\pi_{y|x}\pi_{y'|x'}] - E[\pi_{y|x}]E[\pi_{y'|x'}],
\end{align*}
where the assumed independence between $\theta_{y|x, z}$ and $\theta_{y'|x', z'}$ for $x\neq x'$ and all $z, z'$ implies that
\begin{align*}
E[\pi_{y|x}\pi_{y'|x'}] 
&= \sum_z\sum_{z'}E[\theta_{y|x, z}]E[\theta_{y'|x', z'}]E[\theta_z\theta_{z'}] \\
&= \frac{1}{\alpha+1}\left(\sum_z\sum_{z'}\tilde\alpha_{y|x,z}\tilde\alpha_{y'|x',z}\tilde\alpha_z(\delta_{zz'}+\alpha\tilde\alpha_{z'}) \right)\\
&= \frac{1}{\alpha+1}\left(\sum_z\tilde\alpha_{y|x,z}\tilde\alpha_{y'|x',z}\tilde\alpha_z + \alpha\sum_z\tilde\alpha_{y|x,z}\tilde\alpha_z\sum_{z'}\tilde\alpha_{y'|x',z'}\tilde\alpha_{z'}\right)\\
&= \frac{1}{\alpha+1}\left(\sum_z\tilde\alpha_{y|x,z}\tilde\alpha_{y'|x',z}\tilde\alpha_z + \alpha E[\pi_{y|x}]E[\pi_{y'|x'}]\right).
\end{align*}
Hence, the covariance in this case is given by:
$$
Cov(\pi_{y|x}, \pi_{y'|x'})= \frac{1}{\alpha+1}\left(\sum_z\tilde\alpha_{y|x,z}\tilde\alpha_{y'|x',z}\tilde\alpha_z  -E[\pi_{y|x}]E[\pi_{y'|x'}] \right).
$$

%% file: sections/appendix-illustration-shape-of-posterior.tex
\section{Illustration: The posterior over causal effects}

\begin{figure}[p]
    \centering
    \begin{tabular}{p{.8\linewidth}}\centering
    \begin{tikzpicture}[->, auto,node distance=2.5cm,% 
      main node/.style={font=\sffamily\fontsize{14pt}{10pt}\selectfont,%
      inner sep=2.5mm,outer sep=2.5mm}]
    
      \node[main node] (1) {$X_1$};
      \node[main node] (2) [right of=1] {$X_2$};
      \node[main node] (3) [right of=2] {$X_3$};
      \node[main node] (4) [right of=3] {$X_4$};
    
      \path[]
        (1) edge node [right] {} (2)
        (2) edge node [right] {} (3)
        (4) edge node [right] {} (3)
        (1) edge[bend left] node [left] {} (3);
    \end{tikzpicture}
    \\
    \small (a) A small DAG over four nodes. The edge from $X_1$ into $X_2$ is undirected in the corresponding CPDAG.
    \end{tabular}
    \vspace{1em} \\
    \begin{tabular}{c}
    \includegraphics[width = .75\linewidth]{\rootdir{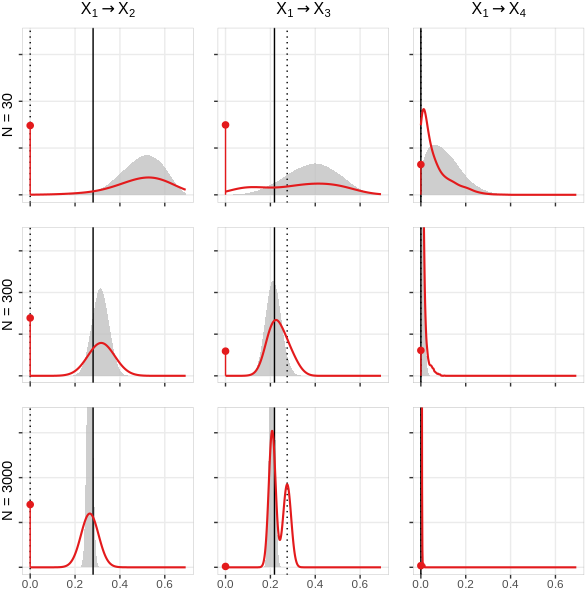}} \\
     \small (b) Posterior distributions over the causal effects of $X_1$ on $X_2, X_3$ and $X_4$, respectively. 
    \end{tabular}
    \caption{Posterior distributions over pairwise causal effects, estimated under the assumption that the DAG is unknown (red) and known (grey) using the back-door formula with parents as adjustment set.
    Under an unknown DAG, the distributions tend to point masses at the set of true causal effects in the two DAGs in the equivalence class. This set is indicated by the black vertical lines, with the solid lines indicating the true DAG. 
    The relative height of the points at zero indicates the probability mass at zero (normalized w.r.t the height of the figure, not the density curves). 
    In the leftmost column, the true effect is non-zero, yet, the equivalence class includes also a zero-effect. 
    In the center column, the effect is non-zero in both DAGs in the equivalence class, but the total effect differs. 
    In the rightmost column, both DAGs imply a zero effect and the posterior distribution tends to a unimodal distribution. 
    Even when the true DAG is assumed known, the effect is typically not set to exactly zero, as the independencies encoded by the DAG are not imposed in the local parent-based estimator.\label{fig::ex_posterior_tau}}
\end{figure}

To illustrate the shape of the posterior and its behavior as the sample size is increased, we generated data from a small toy causal model based on the DAG in Figure \ref{fig::ex_posterior_tau}. We focus on the causal effect that $X_1$ has on the remaining variables, $X_2 , X_3 , X_4$, and estimate the posterior over these effects using parent sets for adjustment. In the CPDAG, representing the Markov equivalence class of the DAG, the edge between $X_1$ and $X_2$ is undirected. Hence, the two DAGs in the class imply different values for the causal effects of $X_1$ on $X_2$ and $X_3$, while the effect of $X_1$ on $X_4$ is zero in both DAGs. Note that the rightmost column in Figure \ref{fig::ex_posterior_tau} also illustrates that, even if the causal DAG is assumed known, the local parent-based estimator does typically not set causal effects to exactly zero for finite samples.

%% file: sections/appendix-sim-setup.tex
\section{Implementation details for the simulation study}
\subsection{Methods}
\subsubsection{BIDA}
We implemented our BIDA procedure in R.\footnote{The current implementation is available at \url{https://github.com/verahk/bida}.} For the computationally more demanding task of identifying the various adjustment sets in a given DAG we used C++. The first step of the procedure is to compute the posterior over the backdoor adjustment sets for each cause-effect pair, as described in Section 3.2 in the paper. 
To this end, we first sampled 1000 DAGs using the hybrid partitionMCMC scheme as implemented in the R-package \texttt{BiDAG} \citep{suterBayesianStructureLearning2023}. We set the maximum number of parents for any node at 13, the default value, and assumed a uniform prior over all DAGs that satisfy this constraint. For the BDeu-prior, we set the equivalent sample size to 1. In the initial screening step, where the PC algorithm is used to learn a start space for the MCMC-procedure, we set the significance level of the conditional independence tests to the default value of $\alpha = 0.05$. 
Next, we identified four different adjustment sets for every cause-effect pair in each of the sampled DAGs, namely: the parent set, the o-set and the minimal versions of both (see Section 2.2 in the paper).
Given the resulting samples of adjustment sets, we proceeded to computing the posterior over $\pi_{ij}$ as described in Section 3.1 in the paper. Also in this case, we set the equivalent sample size of the parameter prior to 1.
We used the posterior mean as the point estimate, both for $\pi_{ij}$ and effects $\tau_{ij}$.

\subsubsection{IDA and optimal-IDA}
We adapted the original IDA \citep{Maathuis2009}  and the optimal IDA \citep{Witte2020} to the categorical setting as follows.
In the structure learning step, we used the stable PC algorithm \citep{colomboOrderIndependentConstraintBasedCausal2014} to infer a CPDAG. Specifically, we used the implementation in the R-package \texttt{pcalg} \citep{Kalisch2012a}. 
 We set the significance level of the conditional independence tests to $\alpha = 0.05$. The sample version of the PC algorithm can result in conflicting edge directions, and the R-package includes different strategies for handling such conflicts. To ensure that the inferred CPDAGs were extendable (as required for the identification of o-sets), we used the option that attempts to redirect ambiguous edges until an extendable structure is found. If no extendable structure is found after 100 edge combinations, the procedure returns a random CPDAG drawn from the inferred skeleton. For each cause-effect pair, we identified all parent sets and all o-sets compatible with the inferred CPDAG, to be used in the original IDA and the optimal-IDA, respectively, and estimated the associated sets of IPTs through the backdoor formula. To this end, we estimated the relevant observational distributions directly using the observed sample counts, adding a small pseudo-count in order to avoid zero probability estimates. We set these pseudo-counts according to the hyperparameter in the BDeu prior, such that the point estimates corresponded to the posterior means of our Bayesian procedure. Finally, we computed causal effect estimates as the JSD of the estimated intervention distributions, and took the average over the set of causal effect estimates to obtain a single point estimate.

\subsubsection{Bestie}
The BMA approach introduced in \cite{moffaUsingDirectedAcyclic2017} is implemented for the binary setting in the R-package \texttt{Bestie}. As the method requires a sample of DAGs as input, we gave in the same sample of DAGs as given to BIDA, and assumed again a BDeu parameter prior with equivalent sample size equal to 1, which is consistent with the parameter prior used to compute the marginal likelihood. 
We used the Monte Carlo based version of the procedure, which approximates the posterior means of the intervention distributions by simulating data from the network under interventions.

\subsubsection{CCDV}
For the BMA approach by \cite{castellettiJointStructureLearning2024}, we used the author's R implementation.\footnote{The implementation is available at \begin{sloppypar}\url{https://github.com/FedeCastelletti/bayes_structure_causal_categorical_graphs}\end{sloppypar}} 
We assumed the same priors as the authors reported for their simulations, namely a Beta(1, 1) for the edge inclusion parameter of the graph prior and an equivalent sample size equal to 1 for the parameter prior. 
The available R implementation targets the average treatment effect on a single outcome variable $Y = X_1$ from all other nodes $X_2, \ldots, X_n$.
To obtain the posterior over the effect for all cause-effect pairs, we iterated over all variables and applied the procedure with each variable in turn defined as the outcome.
In each iteration, we drew 5000 samples and discarded the first 1000 samples as burn-in. In the paper, we report results of a modified version of the procedure in which, conditional on a sampled DAG, the causal effects are computed from the mean values of the (conditional) parameter posterior, rather than a posterior sample of parameters. Appendix D includes results from this modified version and from the original version in which both DAGs and parameters are sampled.

\subsubsection{Naive references}
As naive references, we included unadjusted estimates of the causal parameters given by the corresponding conditional and marginal distributions. More precisely, we estimated each intervention probability $\pi_{x_j|x_i}$ by the conditional probability $\hat\theta_{x_j|x_i}$ as well as the marginal probability $\hat\theta_{x_j}$. Comparing the BIDA and IDA estimates to these two references provides some insight into the sparseness of the inferred structures. For example, given an empty (CP)DAG, both BIDA and IDA return the conditional probabilities when parent sets are used for adjustment. On the other hand, if the o-set is used, they return the marginal probabilities. Also, the conditional probabilities are informative when it comes to assessing the general impact of confounding bias against the bias that arises from incorrectly inferred causal structures.

\subsection{Evaluation criteria}
We use two different evaluation criteria to compare the methods considered: the accuracy of point estimates and the precision to predict strong causal effects. 
 As a measure of accuracy, we use the mean squared error (MSE) between the true values and the point estimates of the target parameters. More specifically, given an $n\times n$ matrix with point-estimates $\hat\tau$ of the causal effects and the true values $\tau$, we used the global MSE (excluding the diagonal elements):
\begin{align*}\label{eq::MSE}
MSE(\hat\tau, \tau) = \frac{1}{n(n-1)}\sum_{i \in V}\sum_{j \in V \setminus \{i\}} (\hat\tau_{ij}-\tau_{ij})^2.
\end{align*}
Similarly, for the matrix $\pi$ with IPTs, we use
\begin{align*}
MSE(\hat\pi, \pi) = \frac{1}{n(n-1)}\sum_{i \in V}\sum_{j \in V\setminus \{i\}}\frac{1}{r_ir_j} \sum_{x_i}\sum_{x_j}(\hat\pi_{x_j|x_i}-\pi_{x_j|x_i})^2.
\end{align*}
From a theoretical point of view, we note that the MSE of the obtained estimates will typically not tend to zero. Instead, due to non-identifiability, the point estimates of BIDA will (ideally) approach the ``average true values'' obtained under the true CPDAG and given the true distribution. These are in general different from the true values obtained under the true DAG.
The two IDA procedures estimate the unique values associated with the inferred CPDAG, yet, not the multiplicity of each value. Hence, at best, the IDA point-estimates tend to the average of the unique effects associated with the true CPDAG, which also in general differ from the true values.

%Unique causal effects can not be estimated from observational data under an unknown DAG, and the estimates of the BIDA procedure can therefore not replace inference based on controlled experiments.
%One of the main motivations behind methods such as BIDA is to inform future experiments by pointing to cause-effect pairs, within a potentially high-dimensional system, that are likely to be important.
%The ability to discover strong causal effects is thus an important property of our method. Low estimation accuracy (in terms of, for example, MSE) is not necessarily a good indicator for this ability. For example, biased estimates of causal effects could still be useful in the context of causal discovery, ranking cause-effect pairs with true large effects higher than those with small or zero effects.

One of the main motivations behind methods such as BIDA is to inform future experiments by pointing to specific cause-effect pairs, out of all possible cause-effect pairs, that are likely to be important.
The ability to discover strong causal effects is thus an important property. High estimation accuracy (in terms of, for example, MSE) is not necessarily a good indicator for this ability. 
For example, biased estimates of causal effects could still be useful in the context of causal discovery, ranking cause-effect pairs with true large effects higher than those with small or zero effects.
To evaluate and compare different rankings of cause-effect pairs, we use the area under the precision-recall curve (AUC-PR).
Given a threshold defining which cause-effect pairs are considered ``strong'', this measure is 1 if a ranking perfectly separates the true and false positives.
For a random ranking, or one with all ties (for example, all effect estimates are zero), the expected AUC-PR equals the rate of true positives.

%% file: sections/appendix-sim-res-binary.tex
\section{Additional simulation results for the binary setting\label{sec::app_res}}

\begin{figure}[ht]
    \centering
   \includegraphics[width=\linewidth]{\rootdir{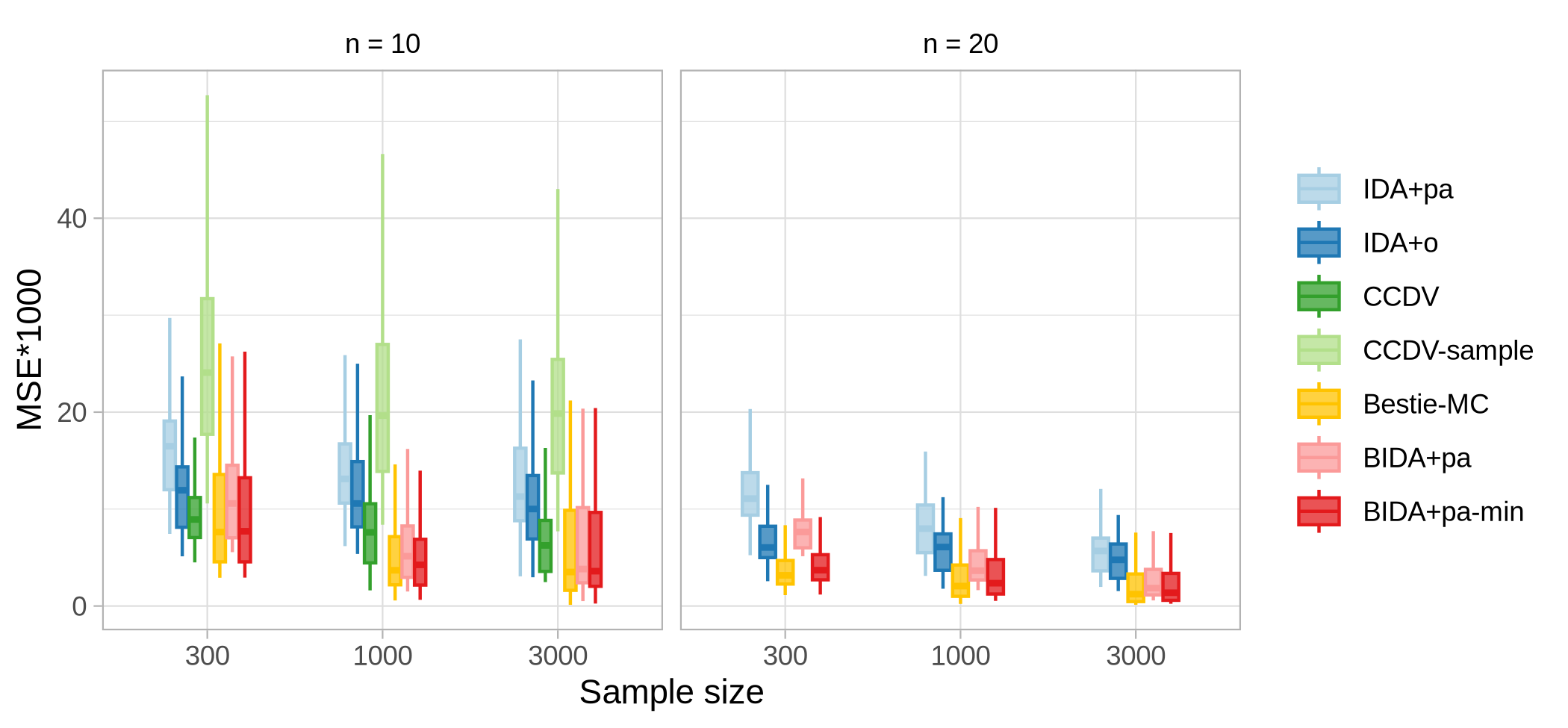}}
    \caption{The accuracy of BIDA with parent sets (BIDA+pa) or minimal parent sets (BIDA+pa-min), as measured by the mean squared errors (MSE) between the true and estimated ATE. Additional methods included are the original IDA (IDA+pa), the optimal IDA (IDA+o), Bestie, CCDV and CCDV's original approach in which the target posterior means are approximated from a joint sample of DAGs and parameters (CCDV-sample). The boxplot shows the distribution over 30 simulation runs, where data sets of different sizes were sampled from a random network of size $n = 10$ (left figure) and $n=20$ (right figure). Outliers are not shown. The CCDV is not included in the $n=20$ setting due to long run times.}
\end{figure}

\begin{figure}[t!]
\centering
\includegraphics[width = \linewidth]{\rootdir{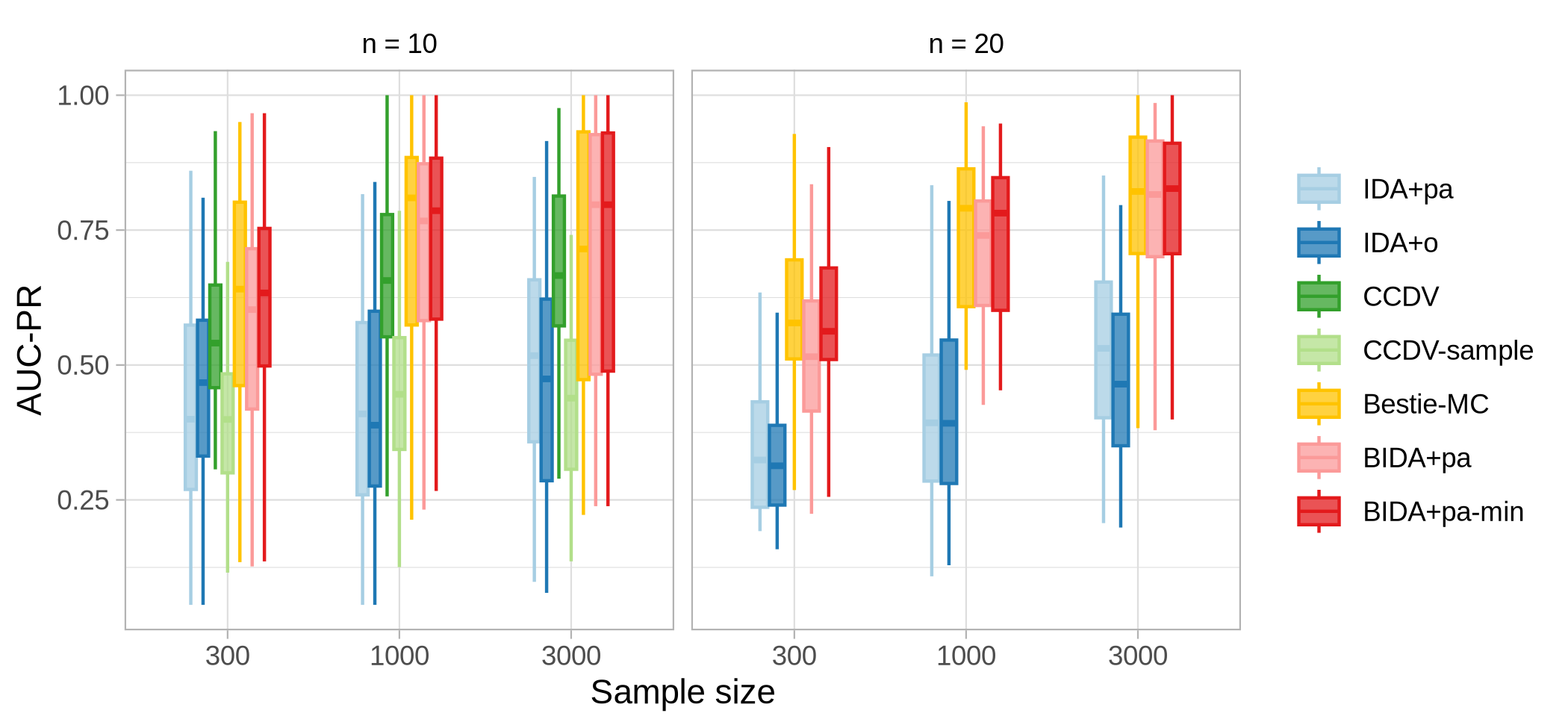}}
\caption{The accuracy of BIDA in predicting strong effects, as measured by the area under the precision-recall curve (AUC-PR), with parent sets (BIDA+pa) and minimal parent sets (BIDA+pa-min). Additional methods included are the original IDA (IDA+pa), the optimal IDA (IDA+o), Bestie, CCDV and CCDV's original approach in which the target posterior means are approximated from a joint sample of DAGs and parameters (CCDV-sample). The true strong effects are defined as the 20 percent largest positive  effects in each network. The boxplot shows the distribution over 30 simulation runs, where data sets of different sizes were sampled from a random network of size $n = 10$ (left figure) and $n=20$ (right figure). Outliers are not shown. The CCDV is not included in the $n=20$ setting due to long run times. \label{fig::sim_binary_aucpr}}
\end{figure}

\begin{figure}[ht]
    \centering
   \includegraphics[width=\linewidth]{\rootdir{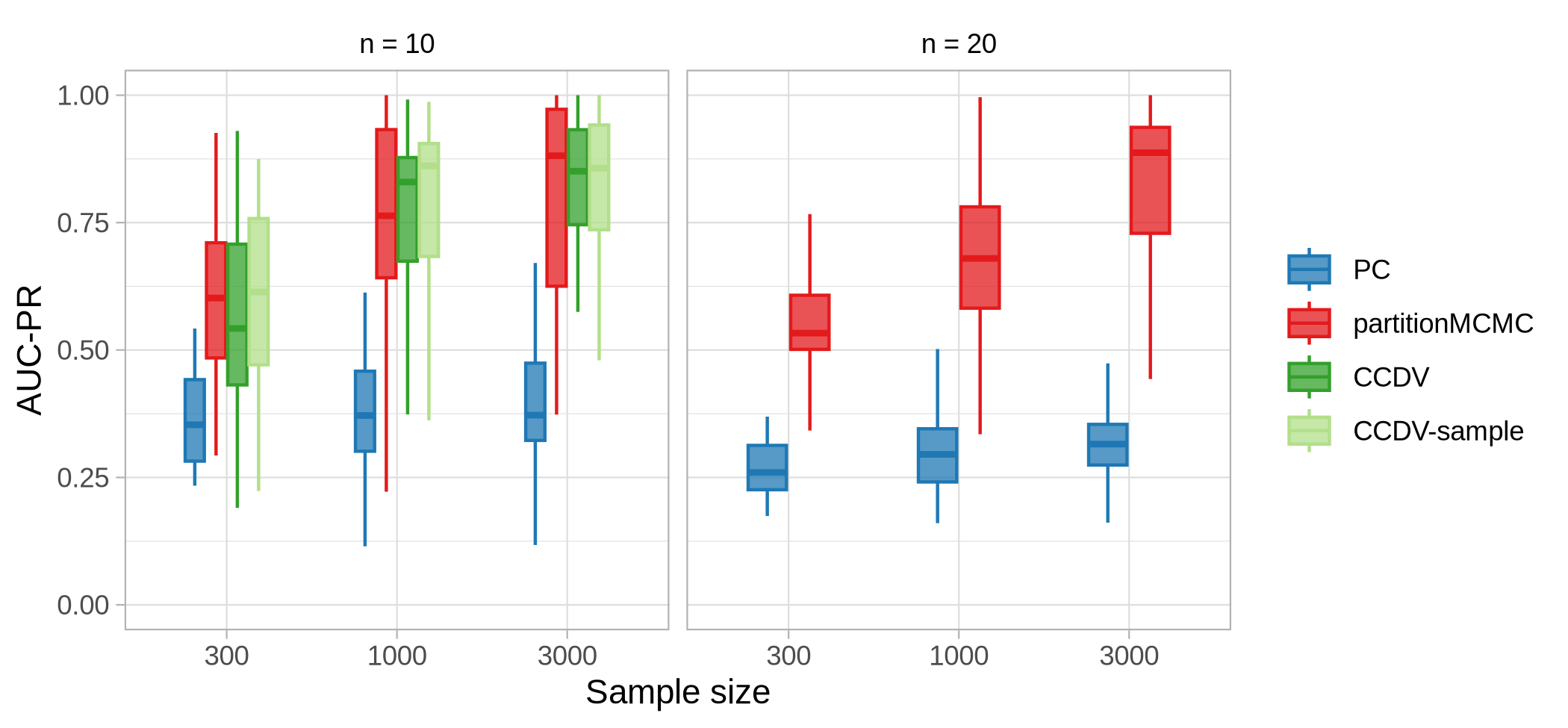}}
    \caption{Average precision in predicting true causal relationships as measured by the AUC-PR, using the ancestor relation probabilities (ARP) of partition-MCMC, the two variants of CCDV and the CPDAG returned by the PC-algorithm. The boxplot shows the distribution over 30 simulation runs. In each run, data sets of different sizes are sampled from a random network of size $n = 10$ to the left and $n=20$ to the right. Outliers are not shown.}
\end{figure}

\input{\rootdir{tables/tab_sim_binary_runtimes}}

%% file: tables/tab_sim_binary_runtimes.tex
\begin{table}[h!]
\caption{Run times (in seconds) averaged over 30 simulation runs, when 6 sessions were run in parallel on a computer with the following CPU: Intel(R) Core(TM) i5-11400 @ 2.60GHz.$^a$ }
\centering
\resizebox{\textwidth}{!}{%
\begin{tabular}{llrrrrrrrrr}
  \toprule\
n & N & {partitionMCMC} & PC & IDA+pa & IDA+o & CCDV & {CCDV-sample} & Bestie-MC & BIDA+pa & BIDA+pa-min \\ 
  \midrule
10 & 300 & 8.07 & 0.04 & 0.13 & 0.14 & 20940.50 & 18211.32 & 19.97 & 0.29 & 0.43 \\ 
  10 & 1000 & 9.73 & 0.13 & 0.11 & 0.12 & 25320.21 & 22305.82 & 9.04 & 0.16 & 0.22 \\ 
   \vspace{1em}10 & 3000 & 8.67 & 0.67 & 0.11 & 0.11 & 30947.39 & 28583.85 & 3.67 & 0.11 & 0.13 \\ 
  20 & 300 & 13.88 & 0.09 & 0.43 & 0.45 & - & - & 107.14 & 1.92 & 3.07 \\ 
  20 & 1000 & 13.19 & 0.38 & 0.48 & 0.50 & - & - & 67.15 & 0.78 & 1.77 \\ 
  20 & 3000 & 14.47 & 1.73 & 0.52 & 0.53 & - & - & 25.23 & 0.50 & 0.76 \\ 
   \bottomrule
\end{tabular}
}
\begin{flushleft}
\noindent$^a${\footnotesize For CCDV, the total time spent sampling DAGs and computing posterior means of all pairwise ATEs is shown. For IDA, Bestie and BIDA only the time spent on parameter inference, conditional on the output for the relevant structure learner (partitionMCMC for Bestie/BIDA and PC for IDA).}
\end{flushleft}
\end{table}

%% file: sections/appendix-sim-res-all.tex
\section{Additional simulation results for the general categorical setting}

\subsection{Results for all 9 benchmark networks}

\begin{landscape}
\begin{figure}[ht]
    \centering
    \includegraphics[width = \linewidth]{\rootdir{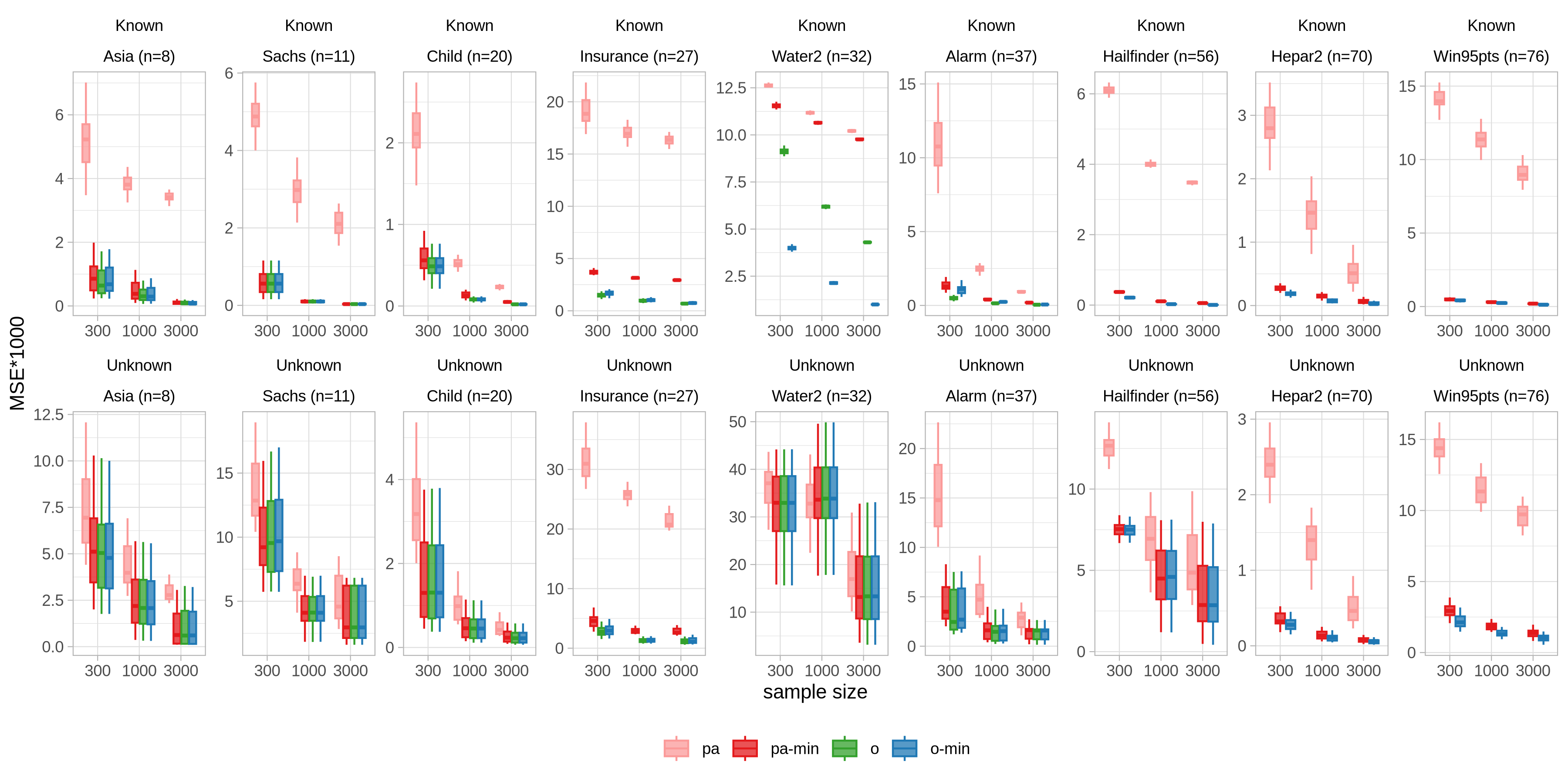}}
    \caption{The accuracy of BIDA using different adjustment sets for adjustment: parent sets (pa), minimal parent sets (pa-min), o-sets (o) or minimal o-sets (o-min), as measured by the mean squared errors (MSE) between the true and estimated causal effects, $\tau$. Each subfigure corresponds to one network and the boxplot shows for each sample size the distribution over 30 independently sampled data sets. Outliers are not shown. In the upper row the DAG is assumed known, in the bottom unknown.\label{fig::mse_adjust_all}}
\end{figure}
\end{landscape}

\begin{figure}[ht]
    \centering
    \includegraphics[width = \textwidth]{\rootdir{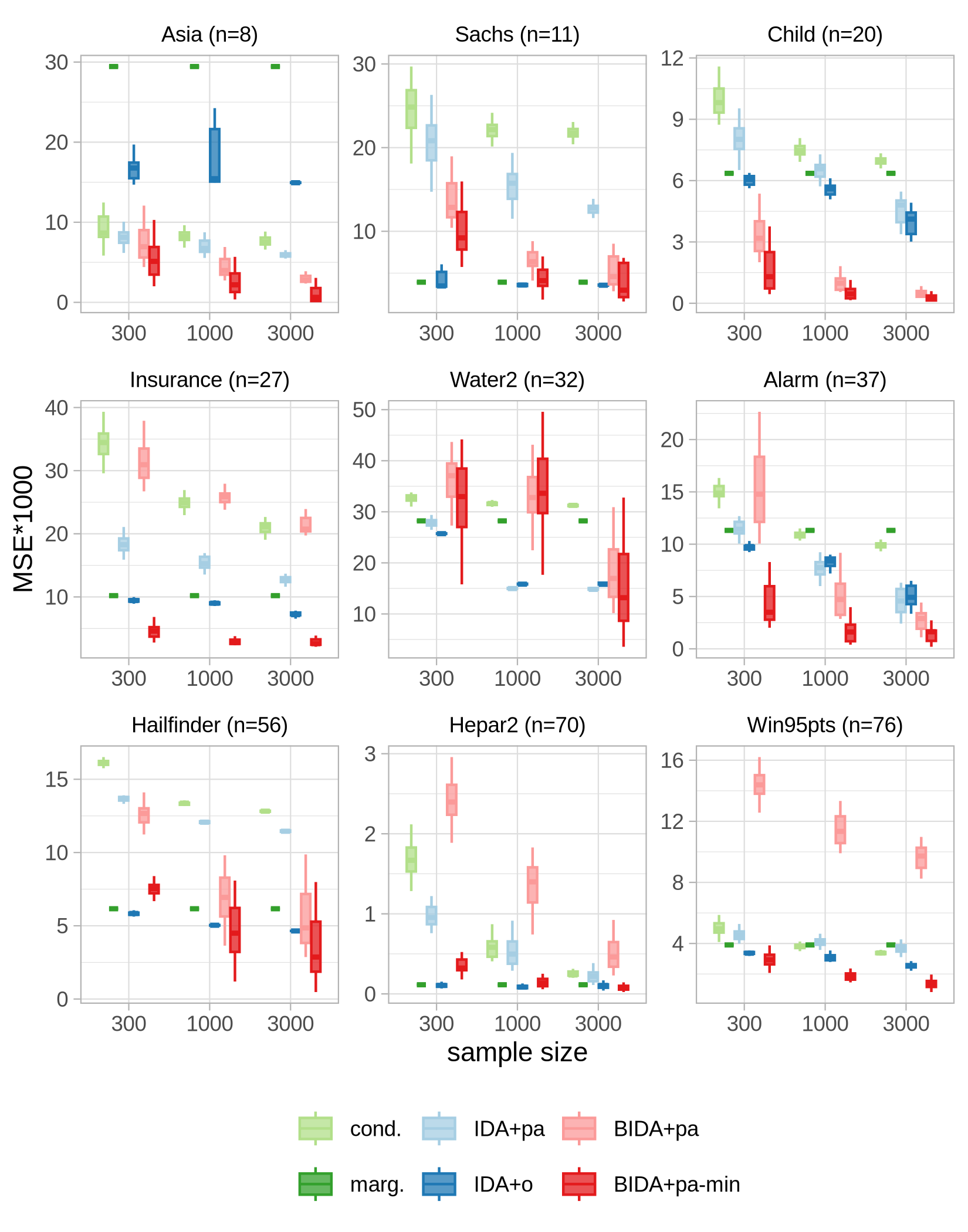}}
    \caption{The accuracy of BIDA using parent (BIDA+pa) or minimal parent sets (BIDA+pa-min) for adjustment, as measured by the mean squared errors (MSE) between the true and estimated causal effects, $\tau$. Additional methods include the marginal probabilities (marg.), conditional probabilities (cond.), the original IDA (IDA+pa) and the optimal IDA (IDA+o). Each subfigure corresponds to one network and the boxplot shows for each sample size the distribution over 30 independently sampled data sets. Outliers are not shown. \label{fig::mse_jsd_all}}
\end{figure}

\begin{figure}[ht]
    \centering
    \includegraphics[width = \textwidth]{\rootdir{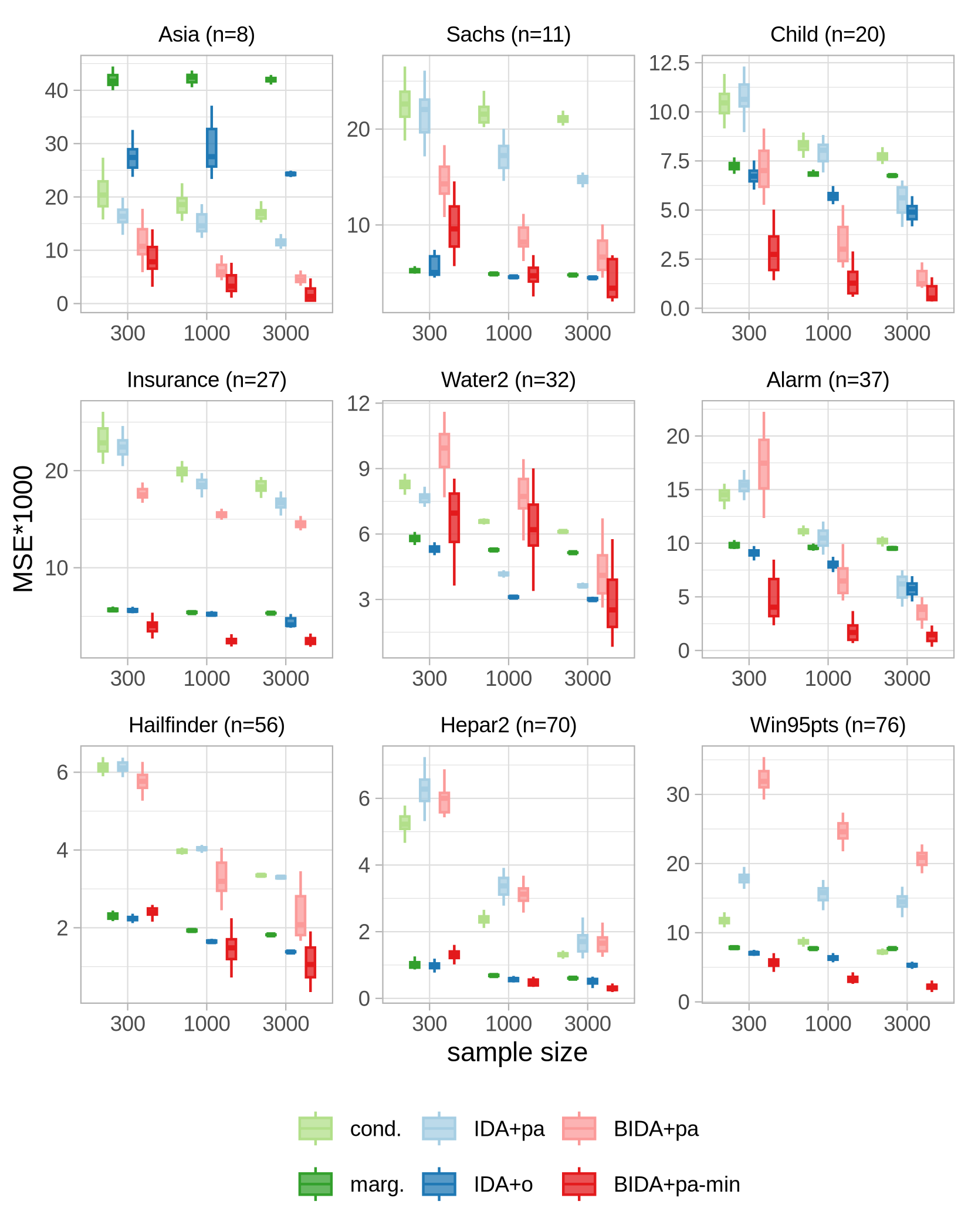}}
    \caption{The accuracy of BIDA using parent (BIDA+pa) or minimal parent sets (BIDA+pa-min) for adjustment, as measured by the mean squared errors (MSE) between the true and estimated intervention probability tables, $\pi$. Additional methods include the marginal probabilities (marg.), conditional probabilities (cond.), the original IDA (IDA+pa) and the optimal IDA (IDA+o). Each subfigure corresponds to one network and the boxplot shows for each sample size the distribution over 30 independently sampled data sets. Outliers are not shown. \label{fig::mse_pdo_all}}
\end{figure}

\begin{figure}[ht]
    \centering
    \includegraphics[width = \textwidth]{\rootdir{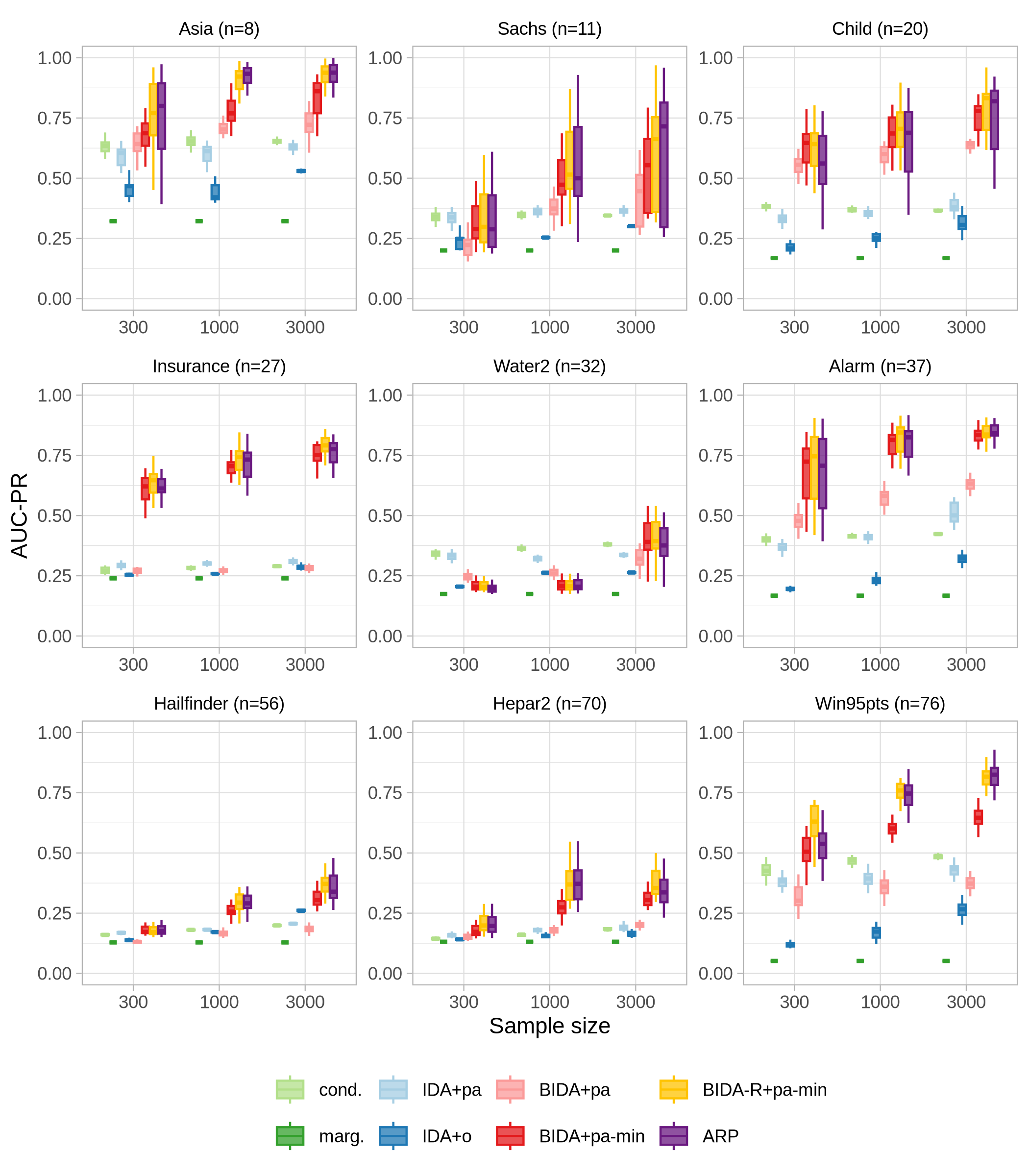}}
    \caption{The accuracy in predicting positive effects, as measured by the area under the precision-recall curve (AUC-PR), using the mean values (BIDA+pa-min) and mean ranks (BIDA-R+pa-min) from the BIDA posterior. Additional methods include the estimated marginal probabilities (marg.), conditional probabilities (cond.), the original IDA (IDA+pa), the optimal IDA (IDA+o) and the ancestor relation probabilities (ARP). 
    Each subfigure corresponds to one network and the boxplot shows for each sample size the distribution over 30 independently sampled data sets. Outliers are not shown. \label{fig::aucpr_pos_all}}
\end{figure}

\begin{figure}[ht]
    \centering
    \includegraphics[width = \textwidth]{\rootdir{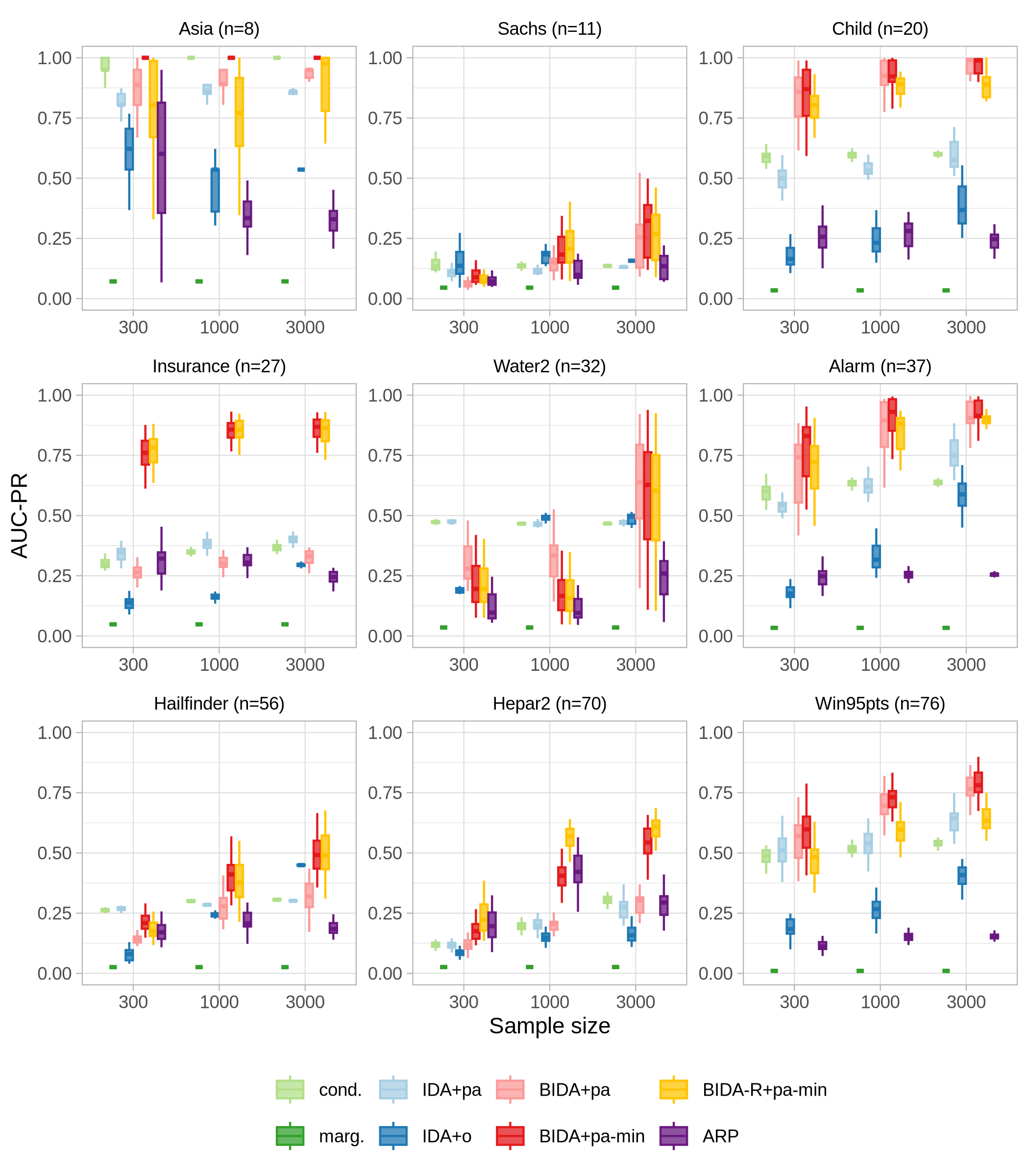}}
    \caption{The accuracy in predicting strong effects (the upper 20 percentile of the non-zero effects), as measured by the area under the precision-recall curve (AUC-PR), using the mean values (BIDA+pa-min) and mean ranks (BIDA-R+pa-min) from the BIDA posterior. Additional methods include the estimated marginal probabilities (marg.), conditional probabilities (cond.), the original IDA (IDA+pa), the optimal IDA (IDA+o) and the ancestor relation probabilities (ARP). 
    Each subfigure corresponds to one network and the boxplot shows for each sample size the distribution over 30 independently sampled data sets. Outliers are not shown.\label{fig::aucpr_top_all}}
\end{figure}

Here we show simulation results for all 9 networks. Characteristics of the networks are shown in Table \ref{tab::networks}
These results include the mean squared error (MSE) of the point-estimates in Figure \ref{fig::mse_adjust_all}-\ref{fig::mse_pdo_all}, estimating the interventional probability tables (IPTs) $\pi$ and the associated causal effects $\tau$ as indicated in the figures, and the area under the precision curve (AUC-PR) attained predicting positive effects (Figure \ref{fig::aucpr_pos_all}) and the strongest effects (Figure \ref{fig::aucpr_top_all}).
Additionally, these plots compare the accuracy of BIDA compared to the two references where we have used the marginal and conditional distributions to estimate the IPTs.

% NETWORK CHARACTERISTICS
\input{\rootdir{tables/network_summary.tex}}

\clearpage

\clearpage 

% ---------------------------------
\subsection{Size of adjustment sets}
Here we compare the size of adjustment sets in the sampled DAGs for three selected networks. 
In each sample of DAGs, for each cause-effect pair and each class of adjustment set we computed:
\begin{itemize}
    \item the number of unique adjustment sets; 
    \item the support of zero-effects;
    \item the maximum adjustment set size;
    \item the average adjustment set size (excluding sets including the effect node).
\end{itemize}
The results are shown in Table \ref{tab::adjsetsize}.
In general, the number of unique adjustment sets tends to decrease with larger sample sizes, indicating that the chain concentrates around some high-scoring node partitionings. 
\input{\rootdir{tables/tab_adjsetsize.tex}}
\clearpage 

% ---------------------------------
\subsection{Run times}
\begin{figure}[ht]
    \centering
    \includegraphics[width = \textwidth ]{\rootdir{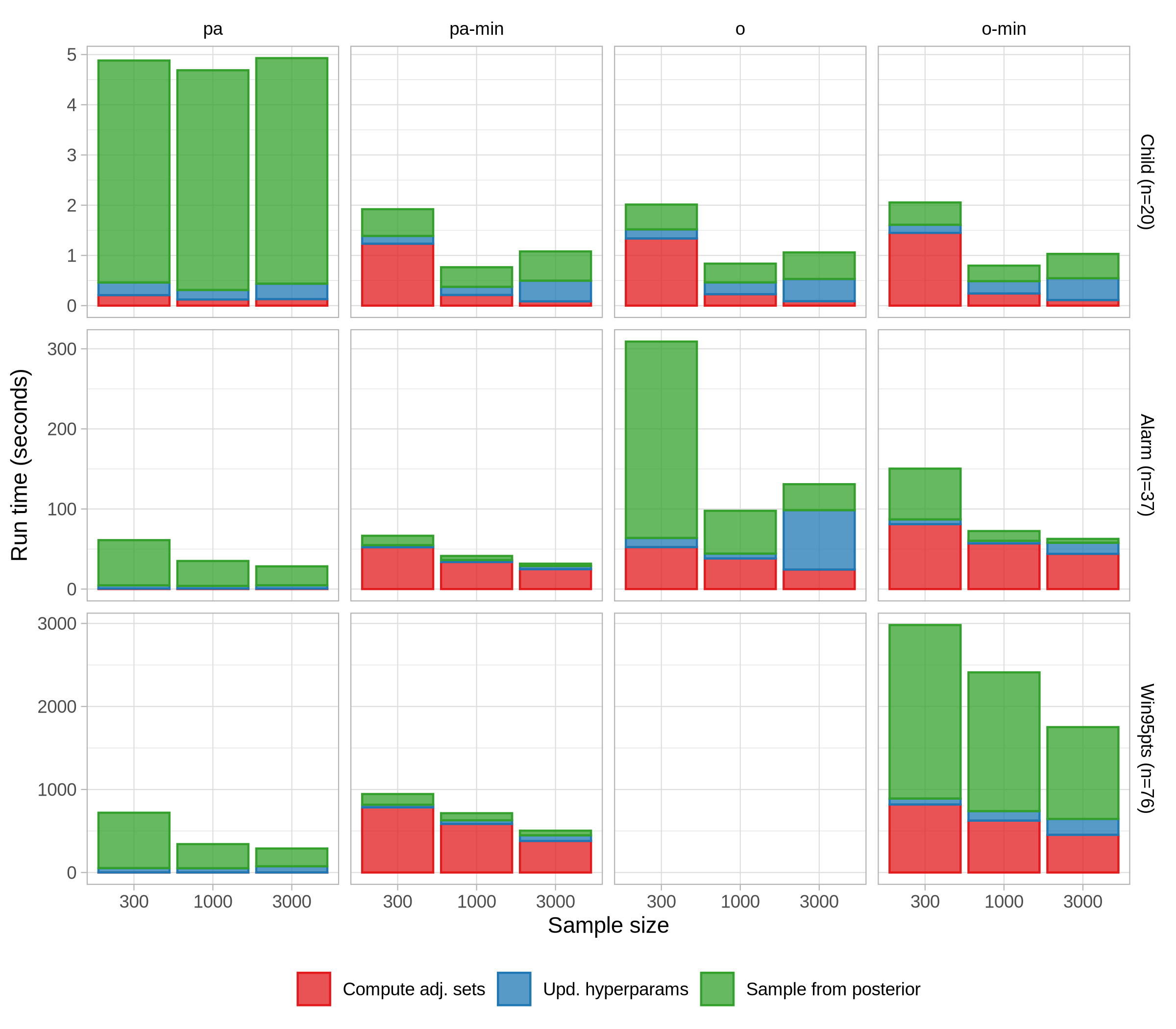}}
    \caption{Average running times of BIDA given a sample of DAGs, decomposed into each step of the estimation procedure. Each column corresponds to a class of adjustment set, and each row to one network. Average over 30 data sets independently sampled from the network, with a sample size of $N = 1000$.  \label{fig::runtimes_decomp}}
\end{figure}

Here, we compare average running times of Bayesian IDA (BIDA) using various adjustment sets when run in parallel on a Intel i5-11400 @ 2.60GHz with 6 cores.
The parent sets (pa) are easy to identify in the sample of DAGs compared to the o-sets (o) and the minimal sets (pa-min, o-min). However, since the parent set implies few zero-effects, larger computational effort is in general required to sample causal effect from the associated posteriors, see Figure \ref{fig::runtimes_decomp}. 
Compared to IDA, BIDA is far less efficient with respect to running time and especially in larger networks. This holds both for the structure learning part of the two procedures, and the inference conditional on the inferred structures (Table D.2). 

\input{\rootdir{tables/tab_runtimes_tot.tex}}

%% file: tables/network_summary.tex
% latex table generated in R 4.2.1 by xtable 1.8-4 package
% Tue May 23 16:50:36 2023
\begin{table}[h]
\centering 
\caption{Characteristics of selected networks from the bnlearn repository. \label{tab::networks}} 
\begingroup
\begin{tabular}{@{}lrrrrrrrrr@{}}
  \hline
    \multirow[c]{2}{*}{Network}
  & \multirow[c]{2}{*}{Size}
  & \multirow[c]{2}{*}{Arcs}
  & \multirow[c]{2}{*}{\shortstack{Anc. \\ rate\textsuperscript{a}}} 
  & \multirow[c]{2}{*}{Params} 
  & \multirow[c]{2}{*}{\shortstack{Markov \\ Blanket}} 
  & \multicolumn{2}{c}{Degree}
  & \multicolumn{2}{c}{Cardinality} \\
    \cline{7-10}
  &
  &
  &
  &
  &
  & Avg. 
  & Max. in  
  & Avg. 
  & Max.  \\ 
  \hline
  Asia       &  8 &   8 &  0.46 &    18 & 2.50 & 2.00 & 2 &  2.00 &  2 \\ 
  Sachs      & 11 &  17 &  0.30 &   178 & 3.09 & 3.09 & 3 &  3.00 &  3 \\ 
  Child      & 20 &  25 &  0.22 &   230 & 3.00 & 2.50 & 2 &  3.00 &  6 \\ 
  Insurance  & 27 &  52 &  0.28 & 1 008 & 5.19 & 3.85 & 3 &  3.30 &  5 \\ 
  Water2\textsuperscript{b}   & 32 &  66 &  0.21 & 10 083 & 7.69 & 4.12 & 5 & 3.62 &  4 \\ 
  Alarm      & 37 &  46 &  0.20 &    509 & 3.51 & 2.49 & 4 & 2.84 &  4 \\ 
  Hailfinder & 56 &  66 &  0.15 &  2 656 & 3.54 & 2.36 & 4 & 3.98 & 11 \\ 
  Hepar2     & 70 & 123 &  0.15 &  1 453 & 4.51 & 3.51 & 6 & 2.31 &  4 \\ 
  Win95pts   & 76 & 112 &  0.06 &    574 & 5.92 & 2.95 & 7 & 2.00 &  2 \\ 
  \hline 
  \multicolumn{10}{l}{\textsuperscript{a}\footnotesize Relative number of ancestor relations, or the share of cause-effect pairs with non-zero effects.}\\[-.25em]
  \multicolumn{10}{l}{\shortstack[l]{\textsuperscript{b}\footnotesize Equivalent to the \emph{Water} network, except for the non-positive marginal distributions over the root \\ \footnotesize nodes,   which are here replaced with uniform distributions.}}\\
\end{tabular} 
\endgroup
\end{table}

%% file: tables/tab_adjsetsize.tex
% latex table generated in R 4.2.1 by xtable 1.8-4 package
% Sun Mar 24 23:40:22 2024
\begin{table}[h]
\centering
\caption{Characteristics of the sampled adjustment sets: the number of unique adjustment sets (No. unique), the rate of adjustment sets that implies no causal effect (prob. zero), and the average and maximum size. Averaged over all cause-effect pairs and 30 simulation runs.} 
\label{tab::adjsetsize}
\begingroup\footnotesize
\begin{tabular}{lllrrrr}
  \toprule
network & N & adjset & No.unique & Prob. zero & Avg. size & Max size \\ 
  \midrule
Child (n=20) & 300 & pa & 3.96 & 0.05 & 1.01 & 3.00 \\ 
  Child (n=20) & 300 & pa-min & 1.52 & 0.84 & 0.30 & 3.00 \\ 
  Child (n=20) & 300 & o & 2.18 & 0.84 & 0.39 & 4.00 \\ 
  Child (n=20) & 300 & o-min & 1.96 & 0.84 & 0.33 & 4.00 \\ 
  Child (n=20) & 1000 & pa & 2.04 & 0.06 & 1.08 & 2.00 \\ 
  Child (n=20) & 1000 & pa-min & 1.19 & 0.84 & 0.23 & 2.00 \\ 
  Child (n=20) & 1000 & o & 1.34 & 0.84 & 0.25 & 3.00 \\ 
  Child (n=20) & 1000 & o-min & 1.32 & 0.84 & 0.23 & 2.00 \\ 
  Child (n=20) & 3000 & pa & 1.60 & 0.06 & 1.21 & 2.00 \\ 
  Child (n=20) & 3000 & pa-min & 1.23 & 0.84 & 0.28 & 2.00 \\ 
  Child (n=20) & 3000 & o & 1.29 & 0.84 & 0.29 & 3.00 \\ 
   \vspace{1em}Child (n=20) & 3000 & o-min & 1.28 & 0.84 & 0.28 & 2.00 \\ 
  Alarm (n=37) & 300 & pa & 19.38 & 0.04 & 1.58 & 7.00 \\ 
  Alarm (n=37) & 300 & pa-min & 10.73 & 0.77 & 0.78 & 5.00 \\ 
  Alarm (n=37) & 300 & o & 55.66 & 0.77 & 2.14 & 12.00 \\ 
  Alarm (n=37) & 300 & o-min & 29.37 & 0.77 & 0.98 & 10.00 \\ 
  Alarm (n=37) & 1000 & pa & 10.46 & 0.04 & 1.41 & 4.00 \\ 
  Alarm (n=37) & 1000 & pa-min & 6.10 & 0.82 & 0.42 & 4.00 \\ 
  Alarm (n=37) & 1000 & o & 18.97 & 0.82 & 1.67 & 9.00 \\ 
  Alarm (n=37) & 1000 & o-min & 9.40 & 0.82 & 0.49 & 7.00 \\ 
  Alarm (n=37) & 3000 & pa & 6.45 & 0.04 & 1.32 & 3.00 \\ 
  Alarm (n=37) & 3000 & pa-min & 3.45 & 0.84 & 0.27 & 3.00 \\ 
  Alarm (n=37) & 3000 & o & 8.66 & 0.84 & 1.71 & 9.00 \\ 
   \vspace{1em}Alarm (n=37) & 3000 & o-min & 4.30 & 0.84 & 0.30 & 6.00 \\ 
  Win95pts (n=76) & 300 & pa & 150.20 & 0.04 & 2.83 & 13.00 \\ 
  Win95pts (n=76) & 300 & pa-min & 78.16 & 0.68 & 2.13 & 13.00 \\ 
  Win95pts (n=76) & 300 & o-min & 185.93 & 0.68 & 4.80 & 38.00 \\ 
  Win95pts (n=76) & 1000 & pa & 118.71 & 0.04 & 2.63 & 10.00 \\ 
  Win95pts (n=76) & 1000 & pa-min & 62.37 & 0.71 & 1.88 & 10.00 \\ 
  Win95pts (n=76) & 1000 & o-min & 163.81 & 0.71 & 3.94 & 34.00 \\ 
  Win95pts (n=76) & 3000 & pa & 98.24 & 0.03 & 2.47 & 9.00 \\ 
  Win95pts (n=76) & 3000 & pa-min & 47.98 & 0.75 & 1.61 & 9.00 \\ 
  Win95pts (n=76) & 3000 & o-min & 126.23 & 0.75 & 3.04 & 32.00 \\ 
   \bottomrule
\end{tabular}
\endgroup
\end{table}

%% file: tables/tab_runtimes_tot.tex
% latex table generated in R 4.2.1 by xtable 1.8-4 package
% Mon Apr 22 12:02:11 2024
\begin{table}[ht]
\centering
\caption{Runtimes (in seconds) for BIDA and IDA using different adjustment sets. Includes the time spent identifying all unique adjustment set and compute point-estimates of the causal effect for all cause-effect pairs, given a sample of DAGs and a inferred CPDAG, respectively. Averaged over 30 simulation runs.} 
\label{tab::runtimes_tot}
\begingroup\footnotesize
\begin{tabular}{lllrrrrr}
  \toprule
&&&& \multicolumn{4}{c}{Point estimates} \\ \cline{5-8}
Method & Network & N & Struct. learning$^a$ & pa & pa-min & o & o-min \\ 
  \midrule
BIDA & Child (n=20) & 300 & 11.48 & 1.84 & 0.65 & 0.68 & 0.74 \\ 
  BIDA & Child (n=20) & 1000 & 11.80 & 1.59 & 0.28 & 0.29 & 0.30 \\ 
   \vspace{1em}BIDA & Child (n=20) & 3000 & 13.28 & 1.77 & 0.34 & 0.34 & 0.34 \\ 
  BIDA & Alarm (n=37) & 300 & 36.77 & 20.24 & 19.82 & 97.78 & 44.56 \\ 
  BIDA & Alarm (n=37) & 1000 & 37.72 & 11.46 & 12.28 & 31.07 & 20.55 \\ 
  \vspace{1em}BIDA & Alarm (n=37) & 3000 & 38.08 & 9.09 & 9.65 & 22.91 & 16.26 \\ 
  BIDA & Win95pts (n=76) & 300 & 758.77 & 545.06 & 941.13 & . & 2992.00 \\ 
  BIDA & Win95pts (n=76) & 1000 & 339.16 & 366.49 & 704.05 & . & 2505.08 \\ 
  \vspace{1em} BIDA & Win95pts (n=76) & 3000 & 298.83 & 309.40 & 512.92 & . & 1798.16 \\ 
  IDA & Child (n=20) & 300 & 0.34 & 0.75 & . & 0.80 & . \\ 
  IDA & Child (n=20) & 1000 & 1.81 & 0.84 & . & 0.84 & . \\ 
   \vspace{1em}IDA & Child (n=20) & 3000 & 10.44 & 0.98 & . & 0.97 & . \\ 
  IDA & Alarm (n=37) & 300 & 1.09 & 2.48 & . & 2.68 & . \\ 
  IDA & Alarm (n=37) & 1000 & 4.29 & 3.40 & . & 3.55 & . \\ 
   \vspace{1em}IDA & Alarm (n=37) & 3000 & 15.95 & 2.94 & . & 3.05 & . \\ 
  IDA & Win95pts (n=76) & 300 & 7.77 & 14.59 & . & 18.72 & . \\ 
  IDA & Win95pts (n=76) & 1000 & 12.55 & 15.66 & . & 20.13 & . \\ 
  IDA & Win95pts (n=76) & 3000 & 53.71 & 14.09 & . & 18.10 & . \\ 
   \bottomrule
\multicolumn{8}{l}{\scriptsize$^a$Run time of partition MCMC and the PC-algorithm for BIDA and IDA, respectively.}
\end{tabular}
\endgroup
\end{table}